\documentclass[12pt, a4paper]{article} 
\usepackage[utf8]{inputenc}

\usepackage{amsmath}
\usepackage{graphicx}
\usepackage{amsfonts}
\usepackage{amssymb}
\usepackage{epsfig}
\usepackage{color}
\usepackage{psfrag}
\usepackage{epstopdf}

\usepackage{caption}
\usepackage{subcaption}

\graphicspath{{figures/}}
\usepackage{xcolor}
\usepackage{epstopdf}

\newcommand{\EQ}{\begin{equation}}
\newcommand{\EN}{\end{equation}}
\newcommand{\be}{\begin{equation}}
\newcommand{\ee}{\end{equation}}
\newcommand{\bea}{\begin{eqnarray}}
\newcommand{\eea}{\end{eqnarray}}

\begin{document} \setcounter{page}{0}
\topmargin 0pt
\oddsidemargin 5mm
\renewcommand{\thefootnote}{\arabic{footnote}}
\newpage
\setcounter{page}{0}
\topmargin 0pt
\oddsidemargin 5mm
\renewcommand{\thefootnote}{\arabic{footnote}}
\newpage

\begin{titlepage}
\begin{flushright}
\end{flushright}
\vspace{0.5cm}
\begin{center}
{\large {\bf Critical points in coupled Potts models and correlated percolation}}

\vspace{1.8cm}
{\large Noel Lamsen$^1$, Youness Diouane$^{2,3}$ and Gesualdo Delfino$^{2}$}\\
\vspace{0.5cm}
{\em $^1$National Institute of Physics, University of the Philippines Diliman,\\ 1101 Quezon City, Philippines}\\
{\em $^{2}$SISSA and INFN -- Via Bonomea 265, 34136 Trieste, Italy}\\
{\em $^{3}$ICTP, Strada Costiera 11, 34151 Trieste, Italy}\\

\end{center}
\vspace{1.2cm}

\renewcommand{\thefootnote}{\arabic{footnote}}
\setcounter{footnote}{0}

\begin{abstract}
\noindent
We use scale invariant scattering theory to exactly determine the renormalization group fixed points of a $q$-state Potts model coupled to an $r$-state Potts model in two dimensions. For integer values of $q$ and $r$ the fixed point equations are very constraining and show in particular that scale invariance in coupled Potts ferromagnets is limited to the Ashkin-Teller case ($q=r=2$). Since our results extend to continuous values of the number of states, we can access the limit $r\to 1$ corresponding to correlated percolation, and show that the critical properties of Potts spin clusters cannot in general be obtained from those of Fortuin-Kasteleyn clusters by analytical continuation. 
\end{abstract}

\end{titlepage}

\tableofcontents
\section{Introduction}
The idea that ferromagnetic transitions correspond to the percolation of clusters of like spins has been present since the early days of the theory of critical phenomena \cite{Fisher} (see \cite{SA} for a review). However, numerical studies for the three-dimensional Ising model \cite{M-K} showed that the natural clusters obtained drawing a link between nearest neighboring spins with the same sign -- we simply call them {\it spin clusters} -- do not percolate at the critical temperature $T_c$ of the magnetic transition. The picture of the ferromagnetic transition as a percolative transition was rescued in \cite{CK}, where it was observed that, in any dimension, a different type of clusters -- the Fortuin-Kasteleyn (FK) clusters \cite{FK} obtained drawing the link between nearest neighboring like spins with a probability determined by the Ising coupling -- do percolate at $T_c$. The FK clusters also satisfy the requirement that their fractal dimension is determined by the scaling dimension of the magnetic order parameter, and allow the coincidence of percolative and magnetic critical exponents. 

A particularly interesting picture emerged in two dimensions, where it was shown that also the Ising spin clusters percolate at $T_c$ \cite{CNPR}, with a new fractal dimension \cite{SG}. Hence FK and spin clusters yield, at the same Ising temperature, two different universality classes of {\it correlated} percolation, which in turn differ from basic ({\it random}) percolation in which there is no interaction among lattice sites \cite{SA}. As random percolation is conveniently brought in the framework of magnetic transitions through its mapping onto the limit $r\to 1$ of the $r$-state Potts model \cite{FK}, Ising-correlated percolation can be described in terms of coupled Ising and $r$-state Potts models, which amount to a dilute Potts model in which the former Ising variables determine if sites are occupied or empty. It is always understood that the auxiliary Potts variables are eventually eliminated by the limit $r\to 1$. The universality classes of FK and spin clusters in the Ising model were then identified in \cite{CK} as corresponding to two different renormalization group (RG) fixed points of this dilute $r\to 1$ Potts model. The fact that the dilute Potts model displays, as  $r$ varies, a critical and a tricritical branch \cite{NBRS,Nienhuis82,Nienhuis87} accomodates for the fixed point of FK clusters on the former and for that of spin clusters on the latter, and led to an exact identification of the fractal dimension of Ising spin clusters \cite{SV}. 

Much insight is usually gained extending to the $q$-state Potts model what has been learned for Ising ($q=2$). The generalization of the above RG picture to $q$-state Potts correlated percolation was studied in \cite{CP}. Now the site variable takes $q$ values, and spin and FK clusters are obtained drawing a link between nearest neighboring sites with the same value, with probability 1 for the former and interaction-dependent for the latter. In \cite{CP} the $q$-state Potts model coupled to the auxiliary $r$-state Potts model was studied by an approximated RG approach, in the relevant limit $r\to 1$. Two fixed points were found as a function of $q$ and were associated to the universality classes of FK and spin clusters. It was conjectured that the two branches coalesce and terminate at the value of $q$ above which the {\it ordinary} Potts transition becomes first order; in two dimensions this value is known to be $q=4$ \cite{Baxter73,Baxter}. This conjecture, however, could never be checked, since the approximate RG of \cite{CP} was unable to see a transition to a first order regime, and the model cannot be numerically simulated in the limit $r\to 1$. The conjecture was extended in \cite{Vanderzande}, where it was proposed that the two branches of fixed points of the coupled $q$-state and $r$-state Potts models can be related, for $r\to 1$, to the critical and tricritical branches of the $q$-state Potts model, which coalesce at $q=4$ and are analytical continuation of each other \cite{Nienhuis82,Nienhuis87}. The idea that the critical properties of spin clusters are related in this way to those of FK clusters was used in \cite{Vanderzande} to propose an exact formula for the fractal dimension of Potts spin clusters as a function of $q$. Given the good agreement of this formula with numerical studies of spin clusters at\footnote{Numerical studies of the Potts model at $q=4$ are notoriously complicated by logarithmic corrections to scaling \cite{CNS}.} $q=3$ \cite{Vanderzande,DBN}, the conjecture about analytic continuation was accepted. 

The three-point connectivity (i.e. the probability that three points are in the same cluster) of $q$-state Potts FK clusters at criticality was exactly determined in \cite{DV_3point}, and was shown to agree with numerical simulations performed for random percolation ($q\to 1$) in \cite{Ziff} and for $q$ generic in \cite{PSVD}. As the first exact analytical result for correlations in critical clusters on the infinite plane after the critical exponents \cite{Nienhuis82}, this connectivity formula also provided a new test for the conjectured analytic continuation for spin clusters. Numerical determination for Potts spin clusters performed in \cite{DPSV} showed the failure of the conjecture for this observable, a finding that reopened the question of the theoretical understanding of spin clusters\footnote{A similar failure of the analytic continuation was then observed in \cite{KEWI} from simulations for the cluster number in geometries with corners, which for FK clusters can be related \cite{Cardy_Peschel} to the Potts central charge \cite{DF}. Tests of conformal invariance for critical Potts spin clusters have recently been performed in \cite{PS}.}.

In this paper we obtain the first exact determination of the RG fixed points for a $q$-state Potts model coupled to an $r$-state Potts model in two dimensions. This is achieved in the framework of scale invariant scattering \cite{paraf} (see \cite{sis} for a review) which in recent years brought new information \cite{random,DT1,DT2,DL_ON1,DL_ON2,ising_vector, DL_softening,DDL_nematic,DLD_RPn,DLD_CPn} on problems which include quenched disorder and liquid crystals. Since our results are obtained for continuous values of $q$ and $r$, we have in particular access to the limit $r\to 1$ relevant for Potts correlated percolation. The subtleties of this limit\footnote{See \cite{DV_3point} for an analytically exact illustration in the context or random percolation.} require that, for a complete description of critical spin clusters, the degrees of freedom of the $q$-state sector and the auxiliary degrees of freedom of the ($r=1+\epsilon$)-state sector are simultaneously critical. We do not find any critical line in the coupled regime along which this requirement is fulfilled with continuity in the whole interval $q\in[2,4]$. One implication is that the conjectured analytical continuation cannot hold in general, thus explaining the failures observed in \cite{DPSV,KEWI}. On the other hand, it may happen, that specific quantities can be evaluated directly {\it at} $r=1$, where the number of $r$-state degrees of freedom is strictly zero and the discontinuities coming from this sector can be ignored. This leaves open the possibility that the formula for the fractal dimension of spin clusters conjectured in \cite{Vanderzande} -- which up to now has been found in agreement with numerical simulations (see also \cite{DPSV}) -- is exact. 

The paper is organized as follows. In the next section we recall how scale invariant scattering theory applies to a single Potts model. In section~\ref{qr} we consider the two coupled models, obtain the fixed point equations, determine their solutions, and discuss some implications of these results. Section~\ref{correlated} is then devoted to the limit $r\to 1$ and its consequences for correlated percolation. A section of concluding remarks and two appendices complete the paper.

\section{$q$-state Potts model}
\label{single}
\subsection{Fixed point equations}
The $q$-state Potts model \cite{Potts,Wu} is defined on the lattice by the reduced Hamiltonian 
\begin{equation}
{\cal H}_q=-J\sum_{\langle i,j \rangle}\delta_{s_i,s_j},
\end{equation}
where $s_i=1,2,\dots,q$ is a "color" variable at site $i$, and the sum is taken over nearest neighbors. The interaction in the form of a Kronecker delta implies that the model possesses the $\mathbb{S}_q$ symmetry under global permutations of the colors. 

\begin{figure}
     \begin{subfigure}[b]{.5\textwidth}
         \centering
         \includegraphics[width=.4\linewidth]{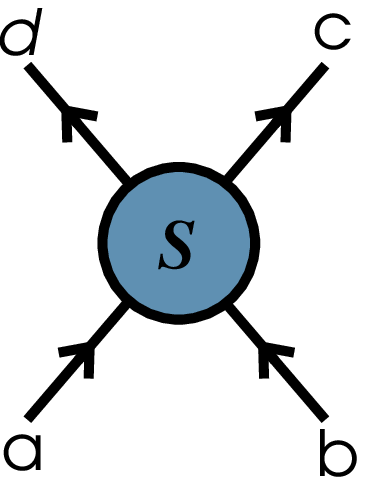}
     \end{subfigure}
     \hfill
     \begin{subfigure}[b]{.5\textwidth}
         \centering
         \includegraphics[width=.3\linewidth]{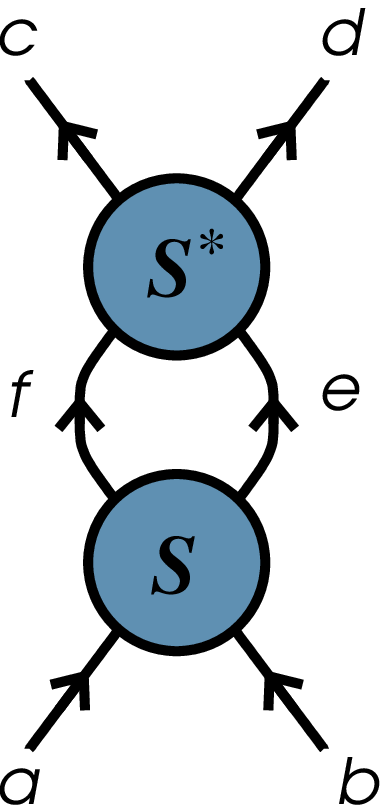}
     \end{subfigure}
\caption{Pictorial representations of the scattering amplitude $S_{ab}^{cd}$ (left) and of the  product of amplitudes entering the unitarity equations~(\ref{unitarity}) (right).  
}
\label{scattering}
\end{figure}

In the two-dimensional case of our interest, the RG fixed points with $\mathbb{S}_q$ symmetry can be obtained, directly in the continuum, within the scale invariant scattering framework \cite{paraf} in which one of the two spatial dimensions plays the role of imaginary time. The method exploits the fact that infinite-dimensional conformal symmetry at fixed points \cite{dFMS} provides infinitely many quantities to be conserved in scattering processes, thus forcing initial and final states to be kinematically identical (complete elasticity). In addition, scale and relativistic invariance lead to energy-independent scattering amplitudes. As a consequence, the equations associated to crossing symmetry, which in relativistic scattering relates amplitudes under exchange of space and time dimensions \cite{ELOP}, and to unitarity, which accounts for conservation of probability, take the particularly simple form \cite{paraf}
\begin{equation}
S_{ab}^{cd}=[S_{ad}^{cb}]^*\,,
\label{cross}
\end{equation} 
\begin{equation}
\sum_{ef} S_{ab}^{ef}[S_{ef}^{cd}]^*=\delta_{ac}\delta_{bd}\,,
\label{unitarity}
\end{equation}
where indices label particle species\footnote{We consider self-conjugated particles.} and our notation for the scattering amplitude $S_{ab}^{cd}$ is illustrated in figure~\ref{scattering}

\begin{figure}[t]
\centering
\includegraphics[scale=1]{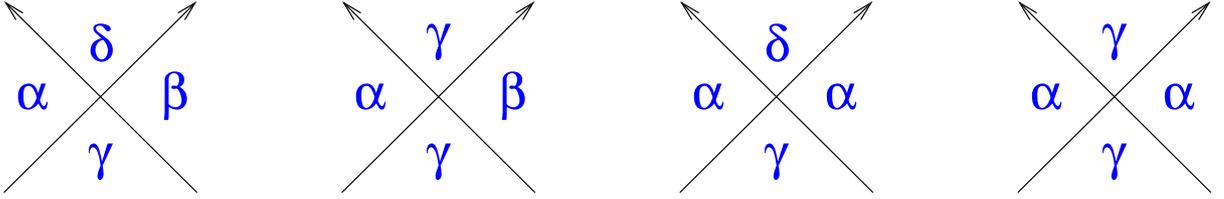}
\caption{Scattering amplitudes $S_0$, $S_1$, $S_2$ and $S_3$ for the Potts model. Different letters correspond to different colors. Time runs upwards.}
\label{qpotts_ampl}
\end{figure}

For the symmety $\mathbb{S}_q$ the scattering theory is implemented considering massless particles that we denote\footnote{In the case of the Potts ferromagnet ($J>0$) below critical temperature, a similar basis of {\it massive} particles corresponds to the kinks that interpolate between pairs of the $q$  degenerate ground states \cite{CZ}. On the lattice, a related representation of the symmetry was used in \cite{BS}.} $A_{\alpha\beta}$, with $\alpha,\beta=1,2,\dots,q$, and $\alpha\neq\beta$. The trajectory of a particle $A_{\alpha\beta}$ is regarded as a line separating a region of two-dimensional space-time characterized by color $\alpha$ from a region characterized by color $\beta$. Permutational symmetry then yields the four inequivalent two-particle scattering amplitudes $S_0$, $S_1$, $S_2$ and $S_3$ shown in figure~\ref{qpotts_ampl}. Then, with the present labeling of the particles, the crossing equations (\ref{cross}) give the relations
\begin{align}
S_0&=S_0^*\equiv\rho_0,\\
S_1&=S_2^*\equiv \rho_1 e^{i \phi},\\
S_3&=S_3^*\equiv \rho_3
\end{align}
where we introduced parametrizations in therms of $\rho_1\geq 0$ and $\rho_0$, $\rho_3$ and $\phi$ real. In this representation the unitarity equations (\ref{unitarity}) take the form \cite{paraf}
\begin{align}
&(q-3)\rho_0^2+\rho_1^2=1,\label{uniq1}\\
&(q-4)\rho_0^2+2\rho_0\rho_1\cos\phi=0, \\
&(q-2)\rho_1^2+\rho_3^2=1,\\
&(q-3)\rho_1^2+2\rho_1\rho_3\cos\phi=0. \label{uniq4}
\end{align} 

Quite relevantly for our purposes, the equations (\ref{uniq1})-(\ref{uniq4}) implement a continuation of the Potts model to noninteger values of $q$. The fact that this analytic continuation is possible is well known on the lattice, where the Potts partition function  allows for the Fortuin-Kasteleyn (FK) expansion \cite{FK}
\EQ
Z_q=\sum_{\{s_i\}}e^{-{\cal H}_q}\propto\sum_{\cal G} p^{N_b}(1-p)^{\bar{N}_b}q^{N_c}\,,
\label{FK}
\EN
where ${\cal G}$ is a graph made of bonds placed on the edges of the lattice, $N_b$ is the number of bonds in ${\cal G}$, $\bar{N}_b$ is the number of edges not covered by a bond, and $N_c$ is the number of clusters in ${\cal G}$, with the convention that a cluster is a set of connected bonds or an isolated site. The quantity 
\EQ
p=1-e^{-J}
\label{prob}
\EN
provides the relation with the coupling $J$ of the spin representation. The FK expansion shows, in particular, that random percolation can be studied as the limit $q\to 1$ of the Potts model. Indeed, in this limit the weight $p^{N_b}(1-p)^{\bar{N}_b}$ of a bond configuration corresponds to random bond occupation with probability $p$. For $q$ generic the FK expansion (\ref{FK}) defines a correlated percolation problem in which each cluster can take $q$ colors and results from bond occupation with the probability (\ref{prob}). These are the FK clusters we referred to in the introduction. Their percolation threshold $p_c$ corresponds to the critical value $J_c$ of the Potts coupling.

In the following we will discuss the Potts model implying the continuation to a noninteger number of colors. An important point, investigated in detail in \cite{DV_4point}, is that, while for the spin model with $q$ integer only the scattering amplitudes of figure~\ref{qpotts_ampl} involving a number of colors not larger than $q$ play a role, all the amplitudes participate to the analytic continuation to $q$ noninteger. The latter case includes random percolation, for which the relevant information is contained in the {\it limit} $q\to 1$. 

\subsection{Solutions}
The solutions of the equations (\ref{uniq1})-(\ref{uniq4}) yield the RG fixed points with $\mathbb{S}_q$ symmetry and are listed in table \ref{qsol}. While we refer the reader to \cite{DT1,sis} for a more detailed discussion of the different solutions, we recall here the points relevant for the remainder of this paper. 

\begin{table}
\begin{center}
\begin{tabular}{|c|c||c|c|c|c|}
\hline
Solution & Range & $\rho_0$ & $\rho_1$ & $2\cos\phi$ & $\rho_3$ \\
\hline
I & $q=3$ &$0$, $2\cos\varphi$ & $1$ & $[-2,2]$ & $0$ \\ 
& & & & & \\
II$_\pm$ & $q\in[-1,3]$ & $0$ & $1$ & $\pm\sqrt{3-q}$ & $\pm \sqrt{3-q}$\\
& & & & & \\
III$_\pm$ & $q\in[0,4]$ & $\pm 1$ & $\sqrt{4-q}$ & $\pm\sqrt{4-q}$ & $\pm (3-q)$\\
& & & & & \\
IV$_\pm$ & $q\in[\frac{1}{2}(7-\sqrt{17}),3]$ & $\pm \sqrt{\frac{q-3}{q^2-5q+5}}$ & $\sqrt{\frac{q-4}{q^2-5q+5}}$ & $\pm\sqrt{(3-q)(4-q)}$ & $\pm \sqrt{\frac{q-3}{q^2-5q+5}}$\\
& & & & & \\
V$_\pm$ & $q\in[4,\frac{1}{2}(7+\sqrt{17})]$ & $\pm \sqrt{\frac{q-3}{q^2-5q+5}}$ & $\sqrt{\frac{q-4}{q^2-5q+5}}$ & $\mp\sqrt{(3-q)(4-q)}$ & $\pm \sqrt{\frac{q-3}{q^2-5q+5}}$\\
\hline
\end{tabular}
\caption{Solutions of Eqs.~(\ref{uniq1})-(\ref{uniq4}). They correspond to RG fixed points with $\mathbb{S}_q$ symmetry.} 
\label{qsol}
\end{center}
\end{table}

The fact that the phase transition of the two-dimensional $q$-state Potts ferromagnet is second order up to $q=4$, and first order above this value \cite{Baxter73}, shows that the ferromagnetic critical line corresponds to solution III of table~\ref{qsol}. In addition, since for $q=2$ (Ising) the only physical amplitude $S_3=\rho_3$ has to take the free fermion value $-1$, solution III$_-$ is further selected. Tricriticality in the Potts model can be realized allowing for vacant sites. The critical line and the tricritical line are known \cite{NBRS} to meet at the endpoint $q=4$, and this is accounted by the fact that solution III$_-$ has two branches differing for the sign of $\sin\phi$. 
 
While for ferromagnets ($J>0$) different lattice structures fall within the same universality class, this is not the case for antiferromagnets ($J<0$), which then require a case by case analysis. The solutions of table~\ref{qsol} follow from $\mathbb{S}_q$ symmetry, independently of the sign of $J$, and should then account also for the RG fixed points of Potts antiferromagnets. Solution I, which is defined only for $q=3$ and contains $\phi$ as free parameter, corresponds to a line of fixed points with central charge $c=1$ \cite{DT1}. One point on the line describes the $q=3$ Potts antiferromagnet on the square lattice, which is known \cite{Baxter,LW,BH,CJS} to possess a $T=0$ Gaussian critical point. More generally, it is expected \cite{DT1} that other points on the line correspond to critical points of $q=3$ antiferromagnets with varying lattice structure. Very interestingly, a family of lattices realizing this phenomenon was recently found \cite{LDJS}. 

On the other hand, the Potts antiferromagnet on the square lattice is known to possess a second order transition for $q\in[0,4]$ \cite{Baxter_square_AF} (with physical values of the temperature only up to $q=3$), and this critical line turns out to be described 
by the scattering solution III$_-$ \cite{DT1}. The Potts critical lines that are presently known to be described by solution III$_-$ are indicated in table~\ref{summary}, where we also specify the central charge and the conformal dimensions of the spin field $s(x)$, of the energy density field $\varepsilon(x)$, and of the field $\eta(x)$ which creates the particles \cite{paraf,DT1}. 
For the ferromagnet these data match those originally obtained on the lattice \cite{Nienhuis82} and identified in the framework of conformal field theory in \cite{DF}; for the square lattice antiferromagnet conformal data were obtained on the lattice in \cite{Saleur} (see also \cite{JS,Ikhlef,DT1}). 

\begin{table}
\begin{center}
\begin{tabular}{c|c|c|c|c|c}
\hline
$\sqrt{q}$ & line & $c$ & $\Delta_{\varepsilon}$ & $\Delta_\eta$ &  $\Delta_s$ \\
\hline
$2\cos\frac{\pi}{p+1}$ & F critical &  $1-\frac{6}{p(p+1)}$ & $\Delta_{2,1}$ & $\Delta_{1,3}$ & $\Delta_{\frac{1}{2},0}$ \\
& & & & &  \\
$2\cos\frac{\pi}{p}$ & F tricritical &  $1-\frac{6}{p(p+1)}$ & $\Delta_{1,2}$ & $\Delta_{3,1}$ & $\Delta_{0,\frac{1}{2}}$ \\
& & & & &  \\
$2\cos\frac{\pi}{N+2}$ & AF square lattice & $\frac{2(N-1)}{N+2}$ & $\frac{N-1}{N}$ &  $\frac{2}{N+2}$ & $\frac{N}{8(N+2)}$ \\
\hline
\end{tabular}
\caption{Realizations of the scattering solution III$_-$ as Potts ferromagnetic (F) and antiferromagnetic (AF) critical lines. $c$ is the central charge and some conformal dimensions are specified using $\Delta_{\mu,\nu}=\frac{[(p+1)\mu-p\nu]^2-1}{4p(p+1)}$.} 
\label{summary}
\end{center}
\end{table}

\begin{figure}
\centering
\includegraphics[width=0.5\textwidth]{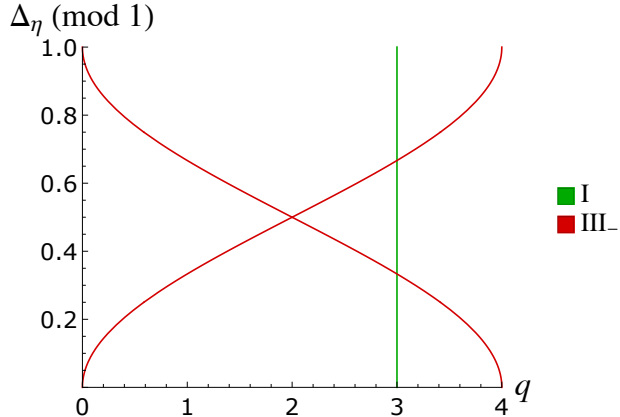}
\caption{Conformal dimension $\Delta_\eta$ determined by equations (\ref{phaseS}) and (\ref{Phi}) for solutions I and III$_-$ of table~\ref{qsol}. \label{delta_plots}}
\end{figure}

Table~\ref{summary} shows that a single scattering solution -- III$_-$ in the present case -- is able to describe different critical lines\footnote{For $q\neq 2$ the scattering amplitudes of the ferromagnet and of the square lattice antiferromagnet become different away from criticality \cite{AF1,AF2}.}. The mechanism through which this happens is most directly illustrated by the conformal dimension $\Delta_\eta$ of the field that creates the particles, which is related by \cite{paraf}
\begin{equation}
\Phi= e^{-2\pi i \Delta_\eta}\,
\label{phaseS}
\end{equation}
to the amplitude
\begin{equation}
\Phi= (q-2)S_{2} + S_{3}\,
\label{Phi}
\end{equation}
through which the state $\sum_{\gamma\neq\alpha} A_{\alpha\gamma}A_{\gamma\alpha}$ scatters into itself. Equation (\ref{phaseS}) shows that scattering theory determines $\Delta_\eta$ mod~1. Solution III$_-$ accounts for three different functions $\Delta_\eta(q)$ for the critical lines of table~\ref{summary}. In the three cases $\Delta_\eta$  changes monotonically as $q$ varies from 0 to 4, going from 0 to 1 for the critical ferromagnetic  branch, from 2 to 1 for the tricritical branch, and from 1 to 0 for the square lattice antiferromagnet (see figure~\ref{delta_plots}). 

Solution V of table~\ref{qsol} has to be noted for the fact that it sets to $(7+\sqrt{17})/2=5.56..$ the maximal value of $q$ for which criticality with $\mathbb{S}_q$ symmetry can be realized in two dimensions\footnote{Quenched disorder brings this maximal value to infinity \cite{AW,HB}, see \cite{random} for the analytical derivation.}, leaving room for a second order transition in a $q=5$ antiferromagnet \cite{DT1}. Numerical evidence in favor of such a transition in the five-state Potts antiferromagnet on the bisected hexagonal lattice was given in \cite{DHJSS}. However, a more recent study concluded in favor of an extremely weak first order transition \cite{Salas_q5}, so that the search for a lattice on which this transition can occur has to start over. Here we notice that the relation $S_0+S_3=S_1+S_2$ among the Potts amplitudes was obtained in \cite{DV_4point} for fixed points possessing an order-disorder duality\footnote{Under this condition the relation extends away from criticality, without the need of integrability \cite{DV_4point}. In particular, it is displayed by the solution for the off-critical ferromagnet, which is integrable \cite{CZ}. Obviously, the relation relies on $\mathbb{S}_q$ symmetry, even if spontaneously broken. See \cite{DG_conf,LTD} for the effects on the spectrum of particle excitations of an explicit breaking of the symmetry induced by a magnetic field.}. Table~\ref{qsol} shows that only the solutions IV and V do not satisfy this relation, thus indicating that they do not allow for order-disorder duality. 

\section{Coupled $q$-state and $r$-state Potts models}
\label{qr}
\subsection{Fixed point equations}
We now want to determine the RG fixed points associated to the symmetry $\mathbb{S}_q\times\mathbb{S}_r$, which can be realized coupling a $q$-state Potts model to an $r$-state Potts model, namely considering the reduced Hamiltonian
\begin{equation}
{\cal H}_{q,r} = -J_1 \sum_{\langle i,j \rangle}  \delta_{s_{i,1}, s_{j,1}}-J_2 \sum_{\langle i,j \rangle}  \delta_{s_{i,2}, s_{j,2}} - J\sum_{\langle i,j \rangle} \delta_{s_{i,1}s_{j,1}}\delta_{s_{i,2}, s_{j,2}}\,, 
\label{H_qr}
\end{equation}
where $s_{i,1} = 1, 2, \ldots, q$ and $s_{i,2}= 1, 2, \ldots, r$ are, respectively, the $q$-state and $r$-state color variables at site $i$. Hence, the index 1 will refer to the $q$-state sector of the coupled model, and index 2 to the $r$-state sector. Since our analysis will be performed directly in the continuum limit and will only rely on symmetry, our results for the critical points will cover all combinations of ferromagnetic and antiferromagnetic values of the couplings $J_1$, $J_2$ and $J$. 

\begin{figure}[t]
\centering
\includegraphics[scale=1.2]{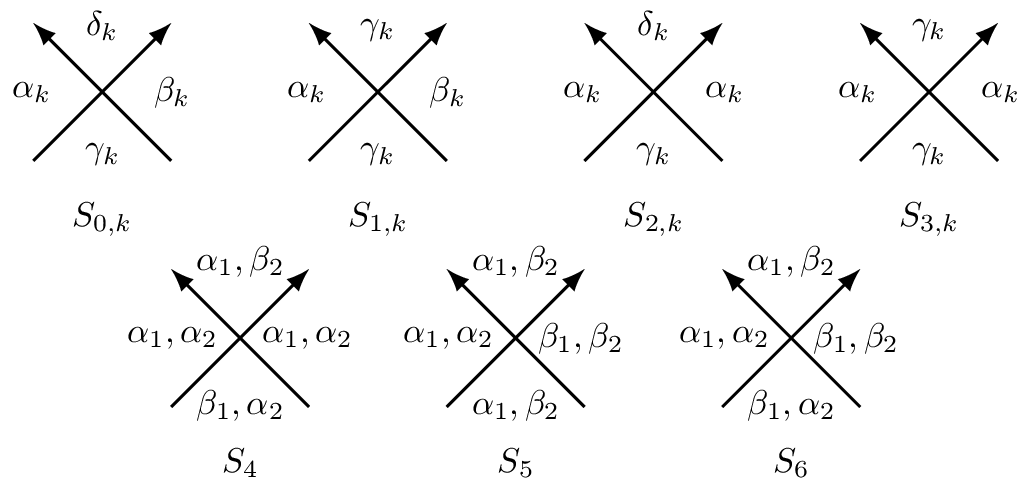}
\caption{Scattering amplitudes for the coupled $q$-state and $r$-state Potts models. The label $\alpha_k$ refers to color $\alpha$ in the $q$-state model for $k=1$, and in the $r$-state model for $k=2$.}
\label{twopotts}
\end{figure}

The implementation of the scattering description follows the steps already seen in the previous section for the basic Potts model. In the first place, there are particle excitations associated to each sector. We denote them $A_{\alpha_k\beta_k}$, where $k=1,2$ labels the two sectors, $\alpha_1,\beta_1=1,\dots, q$ and $\alpha_2,\beta_2=1,\dots, r$ ($\alpha\neq \beta$). Now the trajectory of a particle $A_{\alpha_k\beta_k}$ separates a region of space-time characterized by the colors $\alpha_1$ in the $q$-state sector and $\alpha_2$ in the $r$-state sector from a region in which sector $k$ changes its color to $\beta_k$, with the color of the other sector remaining unchanged. It follows that the two-particle scattering amplitudes inequivalent under color permutations are those\footnote{We also imply that the theory is invariant under time reversal and spatial reflection.} depicted in figure \ref{twopotts}. The first four amplitudes involve only particles belonging to the same sector, so that we keep for them the notation of the previous section, up to the addition of the sector index $k$. On the other hand, the remaining three amplitudes involve particles from both sectors and are responsible for the coupling of the two Potts models.

The crossing symmetry equations (\ref{cross}) now become
\begin{align}
&S_{0,k}=S_{0,k}^*\equiv\rho_{0,k},\\
&S_{1,k}=S_{2,k}^*\equiv\rho_{1,k} e^{i \phi_k},\\
&S_{3,k}=S_{3,k}^*\equiv\rho_{3,k},\\
&S_4=S_5^*\equiv\rho_4 e^{i\theta},\\
&S_6=S_6^*\equiv\rho_6,
\end{align}
with parametrizations in terms of $\rho_{1,k}$ and $\rho_4$ nonnegative, and $\rho_{0,k}$, $\rho_{3,k}$, $\rho_6$, $\phi_k$ and $\theta$ real. The unitarity equations (\ref{unitarity}) then take the form
{\allowdisplaybreaks
\begin{align}
&0=(q-4)\rho_{0,1}^2+2\rho_{1,1}\rho_{0,1}\cos\phi_1,\label{uni1}\\
&0=(r-4)\rho_{0,2}^2+2\rho_{1,2}\rho_{0,2}\cos\phi_2, \label{uni2}\\
&1=(q-3)\rho_{0,1}^2+\rho_{1,1}^2, \label{uni3}\\
&1=(r-3)\rho_{0,2}^2+\rho_{1,2}^2,\label{uni4}\\
&0=(q-3)\rho_{1,1}^2+2\rho_{1,1}\rho_{3,1}\cos\phi_1+(r-1)\rho_4^2,\label{uni5}\\
&0=(r-3)\rho_{1,2}^2+2\rho_{1,2}\rho_{3,2}\cos\phi_2+(q-1)\rho_4^2, \label{uni6}\\
&1=(q-2)\rho_{1,1}^2+\rho_{3,1}^2+(r-1)\rho_4^2,\label{uni7}\\
&1=(r-2)\rho_{1,2}^2+\rho_{3,2}^2+(q-1)\rho_4^2, \label{uni8}\\
&0=\rho_4\left[\rho_{3,2}e^{i\theta}+\rho_{3,1}e^{-i\theta}+(q-2)\rho_{1,1}e^{-i(\theta+\phi_1)}+(r-2)\rho_{1,2}e^{i(\theta+\phi_2)}\right],\label{uni9}\\
&1=\rho_4^2+\rho_6^2, \label{uni10}\\
&0=2\rho_4\rho_6\cos\theta.\label{uni11}
\end{align}
}
Notice that $\rho_4=0$ directly yields $S_4=S_5=0$; in addition (\ref{uni10}) implies $S_6=\pm 1$, namely absence of scattering between particles from different sectors. It follows that $\rho_4=0$ corresponds to the case in which the two Potts models decouple, and indeed the system (\ref{uni1})-(\ref{uni11}) gives back, for each sector, the equations of the previous section.

\subsection{Solutions}
\label{fpsolutions}
The solutions of equations \eqref{uni1}-\eqref{uni11} yield the RG fixed points with $\mathbb{S}_q\times\mathbb{S}_r$ symmetry. It turns out that the space of solutions can be divided into three subspaces according to the values taken by $\rho_4$. The first subspace is that in which the $q$-state sector and the $r$-state sector are decoupled ($\rho_4=0$); in this case the solutions can be immediately traced back to those we saw in the previous section for a single Potts model and do not need further discussion. The second subspace is that in which $\rho_4$ varies; we will refer to the solutions in this subspace as solutions of type V. The third subspace is that of solutions with $\rho_4=1$, which we will call solutions of type S; clearly, in the S-type solutions the two sectors are always strongly coupled. We will also use the notations
\EQ
 x_k=\rho_{1,k}\cos\phi_k\,,\hspace{1cm}y_k=\rho_{1,k}\sin\phi_k\,,\hspace{1.5cm}k= 1, 2\,,
\label{xy}
\EN
while the notation $(\pm)$ will indicate that both signs are allowed. Notice in particular that, due to the form of equations (\ref{cross}) and (\ref{unitarity}), given a solution, another solution is obtained reversing the sign of all amplitudes.

The Hamiltonian (\ref{H_qr}) and the equations \eqref{uni1}-\eqref{uni11} are invariant under the simultaneous exchanges 
\EQ
q\leftrightarrow r\,, \hspace{1cm}\textrm{sector index}\,\,1\leftrightarrow \textrm{sector index}\,\,2\,.
\label{exchange}
\EN
It follows that this exchange operation maps a solution of the fixed point equations into another solution. Some solutions will be mapped into themselves and will be called exchange-invariant solutions. In these invariant solutions the $q$-state sector and the $r$-state sector play a symmetric role: they are both ferromagnetic, or both antiferromagnetic on the same lattice. This is the case normally considered when referring to coupled Potts models, and in this section we will list the exchange-invariant solutions. More generally -- for example for the application to correlated percolation of the next section -- it is relevant to know also the noninvariant solutions, which we then list in appendix~\ref{noninv}. If a noninvariant solution possesses a decoupling limit $\rho_4\to 0$, the amplitudes obtained in the limit for the $q$-state sector and for the $r$-state sector correspond to different solutions\footnote{This was not possible in the case of \cite{random}, where $n$ identical Potts replicas were considered for the purpose of studying quenched disorder ($n\to 0$).} of table~\ref{qsol}.

\subsubsection{Solutions with varying $\rho_4$}
We give here the solutions of type V invariant under the exchange operation (\ref{exchange}). They are denoted by V followed by a number distinguishing the different solutions. These numbers come from a different selection process and are not presented in progressive order. 

\begin{figure}
\centering
\includegraphics[width=0.32\textwidth]{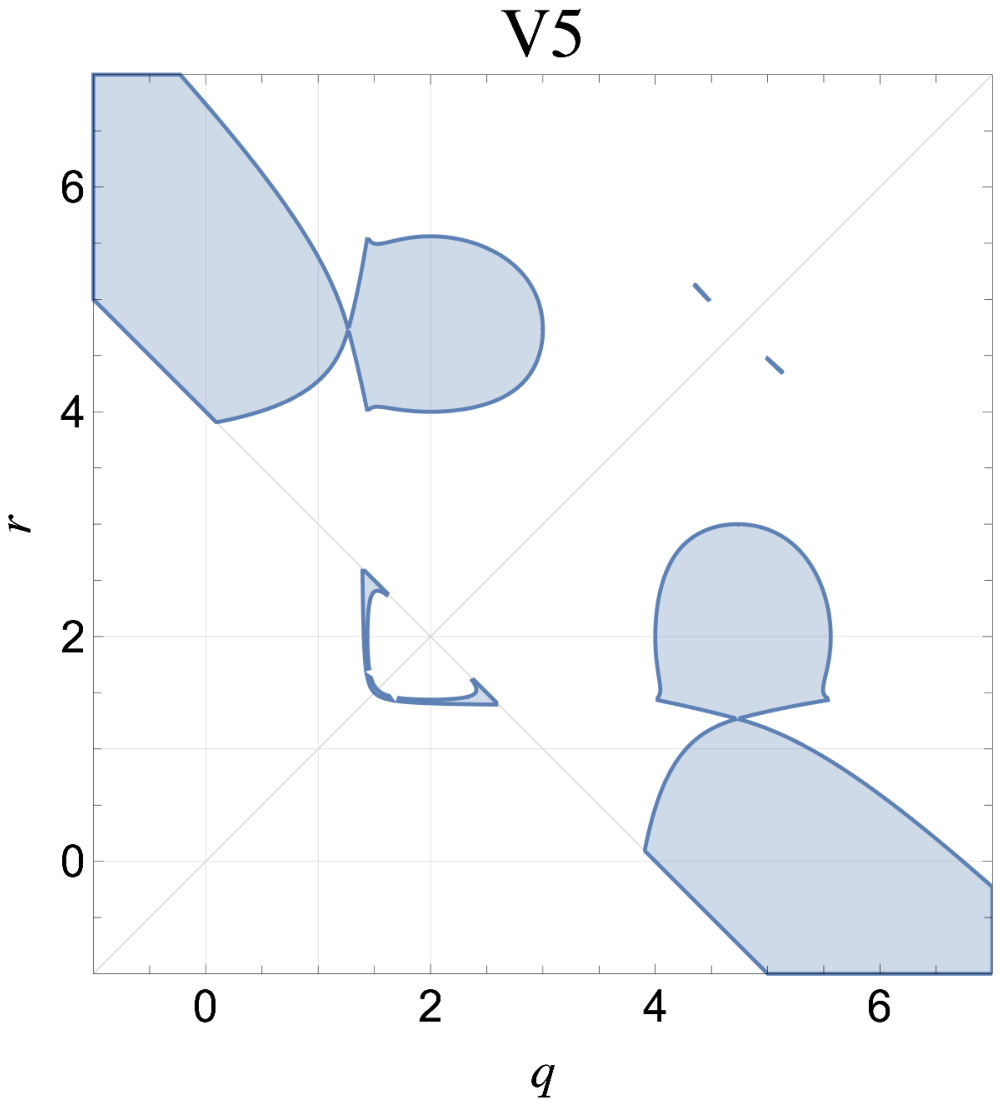}\, 
\includegraphics[width=0.32\textwidth]{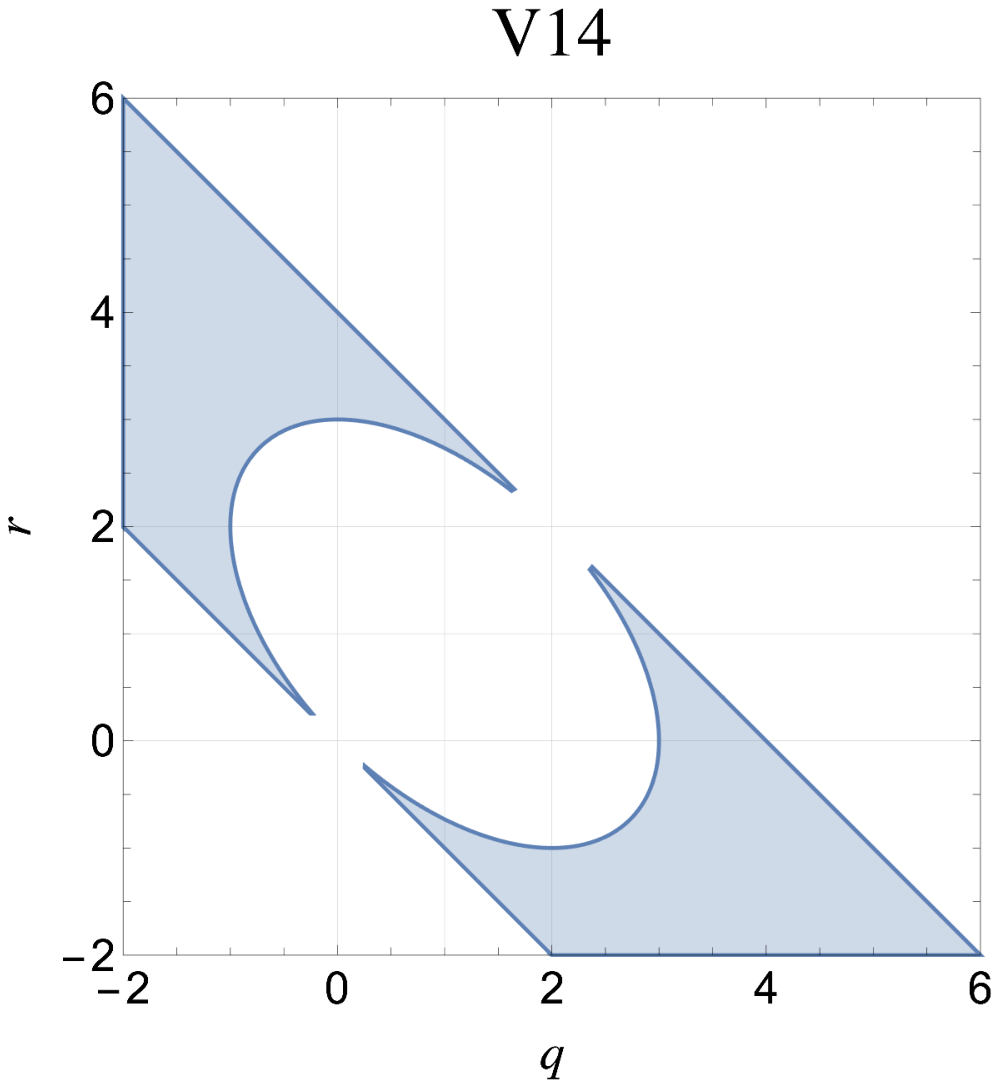}\, 
\includegraphics[width=0.32\textwidth]{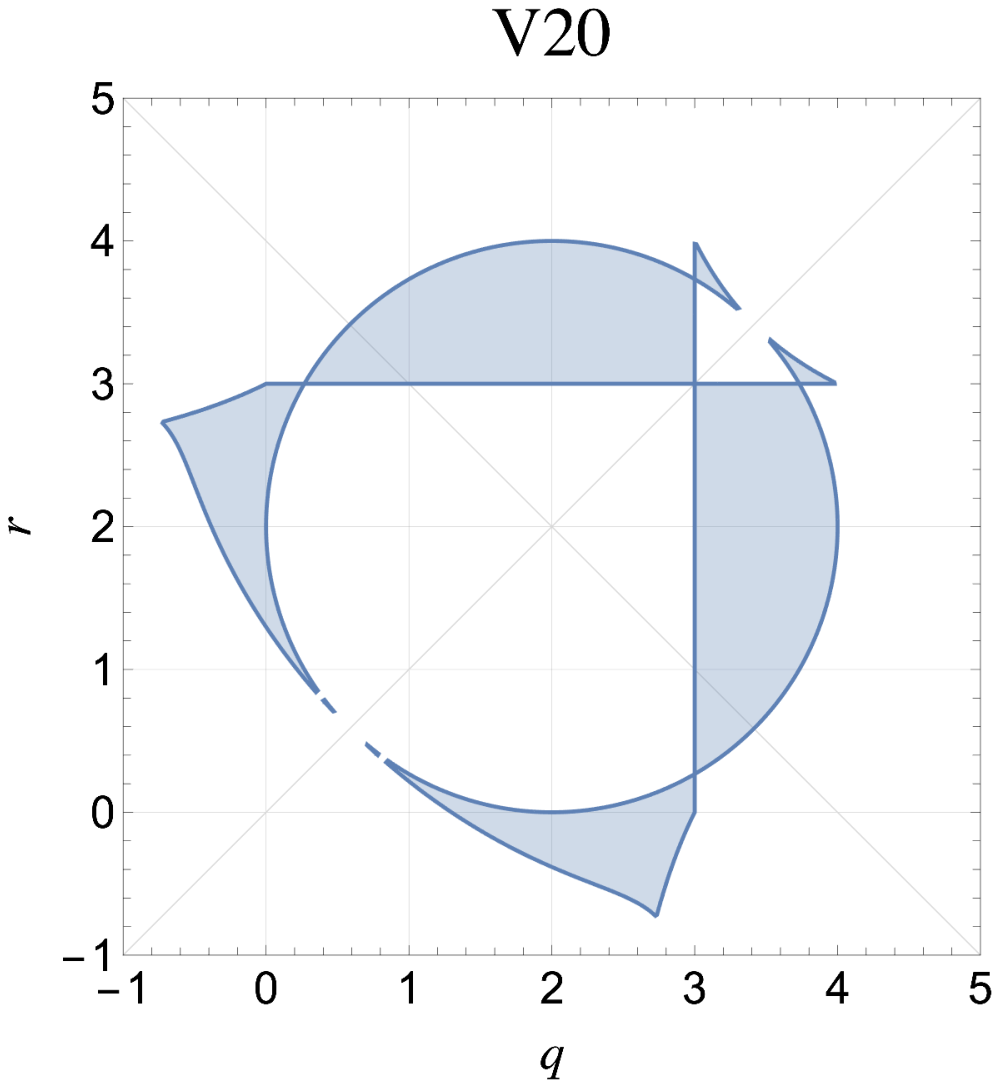}
\caption{Domains of definition in the $q$-$r$ plane for the solutions V5, V14 and V20.
\label{fig:Vsymgen}}
\end{figure}

A first group of solutions -- V5, V14 and V20 -- are defined in the ranges of $q$ and $r$ shown in figure~\ref{fig:Vsymgen}. They read\\
$\bullet$ V5
\begin{align}
\begin{split}
& \rho_{0,1}=\rho_{3,1}=\pm(r^2-6r+6)f(q,r) , \quad  \rho_{0,2}=\rho_{3,2}=\pm(q^2-6q+6)f(q,r) \\ 
& x_1 = -\frac{q-4}{2}\rho_{0,1}, \quad y_1 = (\pm) \frac{r-2}{2}g(q,r), \quad x_2 = -\frac{r-4}{2}\rho_{0,2}, \quad y_2 = -\frac{q-2}{r-2}y_1 \\
&\rho_4=\sqrt{\frac{\frac{(q^2-6 q+6)^2}{q^2-5 q+5}(q-3) -\frac{ (r^2-6 r+6)^2}{r^2-5 r+5}(r-3)}{\frac{(r^2-6 r+6)^2}{r^2-5 r+5}(q-1)-\frac{(q^2-6 q+6)^2}{q^2-5 q+5}(r-1)}}, \quad \theta = (\pm) \frac{\pi}{2}, \quad \rho_6=(\pm)\sqrt{1-\rho_4^2},\\
& f(q,r)=\sqrt{\frac{4-q-r}{(q r-q-r) [(q^2-5 q+5) (r^2-5 r+5) - qr + q + r - 1 ]}},\\
& g(q,r)=\sqrt{\frac{(q r-2 q-2 r+6) [(q^2-6 q+6) (r-2)+(q-2) (r^2-6 r+6)]}{(q r-q-r) [(q^2-5q+5) (r^2-5 r+5) - qr + q + r - 1]}},
\end{split}
\end{align}
$\bullet$ V14
\begin{align}
\begin{split}
&\rho_{0,1}=\rho_{0,2} = 0 , \quad \rho_{3,1}=2x_1 , \quad \rho_{3,2}=2x_2, \quad x_1=\pm \frac{r}{2}\sqrt{\frac{4-q-r}{q+r-qr}} ,  \\
& y_1=(\pm)\frac{r-2}{2}\sqrt{\frac{q+r}{q+r-qr}},  \quad y_2= -\frac{q-2}{r-2}y_1, \quad x_2=\frac{q}{r}x_1 ,\\
&\rho_4=\sqrt{\frac{q^2+r^2 -3(q+r) + q r}{q + r - qr}} \, \, , \, \, \rho_6=(\pm)\sqrt{1- \rho_4^2} , \quad \theta=(\pm)\frac{\pi}{2},
\end{split}
\end{align}
$\bullet$ V20
\begin{align}
\begin{split}
&\rho_{0,1}=\pm\frac{r^2-4 r+2}{\sqrt{1-(q-3) (q-1) (r-3) (r-1)}}, \quad \rho_{0,2} = \frac{q^2 -4q + 2}{r^2-4r+2} \rho_{0, 1}, \\
&x_1= -\frac{q-4}{2}\rho_{0,1}, \quad x_2 = -\frac{r-4}{2} \rho_{0,2}, \quad \rho_{3,1} = -(q-3)\rho_{0,1}, \quad \rho_{3,2} = -(r-3)\rho_{0,2}, \\
&y_1= (\pm)\frac{r-2}{2}\sqrt{\frac{4-(q-2)^2 (r-2)^2}{1-(q-3) (q-1) (r-3) (r-1)}}, \quad y_2=-\frac{q-2}{r-2}y_1,\\
&\rho_4=\sqrt{\frac{\left(q^2-4 q+2\right)^2-\left(r^2-4 r+2\right)^2}{\frac{q-1}{r-3}(r^2-4 r+2)^2-\frac{r-1}{q-3}(q^2-4 q+2)^2}},\quad \theta = (\pm) \frac{\pi}{2},  \quad \rho_6=(\pm)\sqrt{1-\rho_4^2}.
\end{split}
\end{align}

\begin{figure}[t]
\centering
\includegraphics[width=0.45\textwidth]{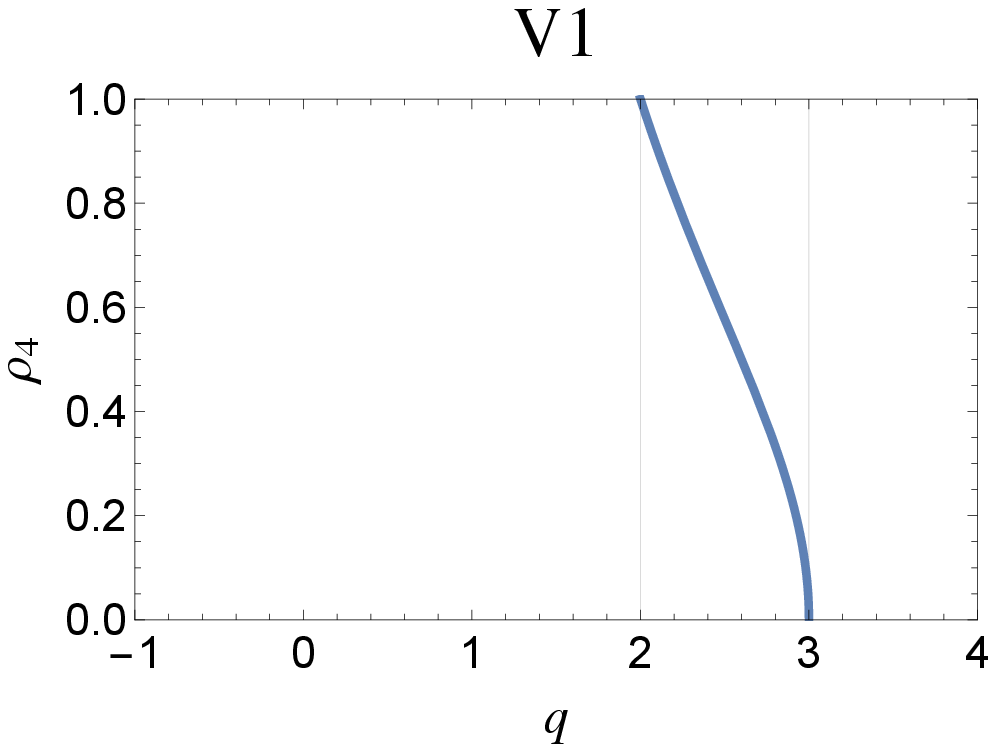}\hspace{25pt}
\includegraphics[width=0.45\textwidth]{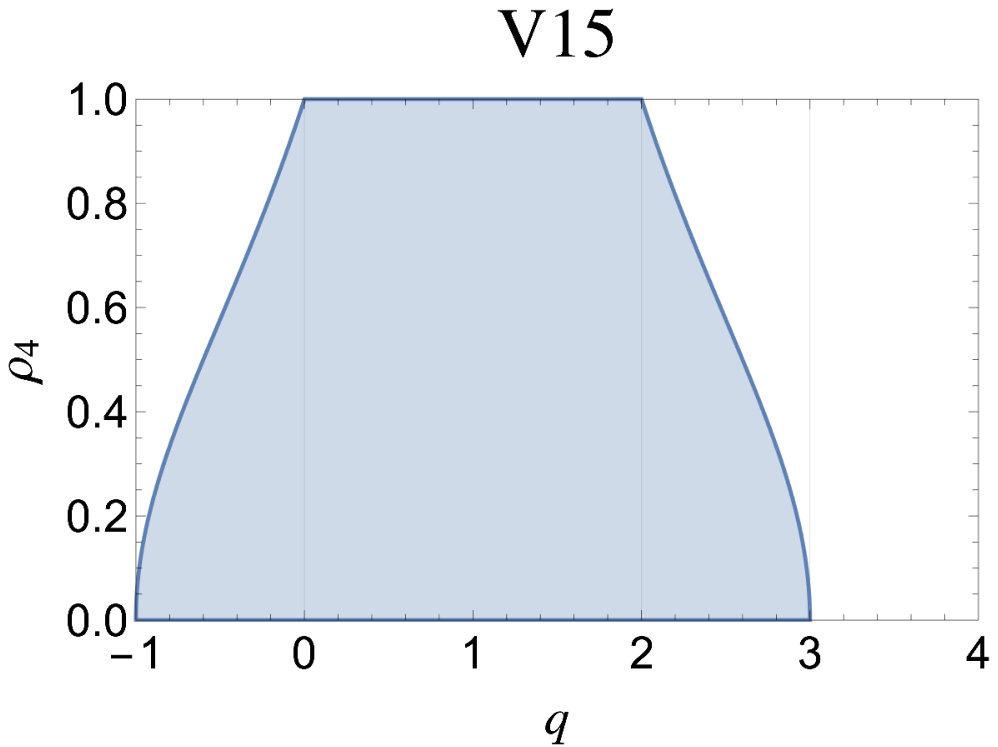} \\[3ex]
\includegraphics[width=0.45\textwidth]{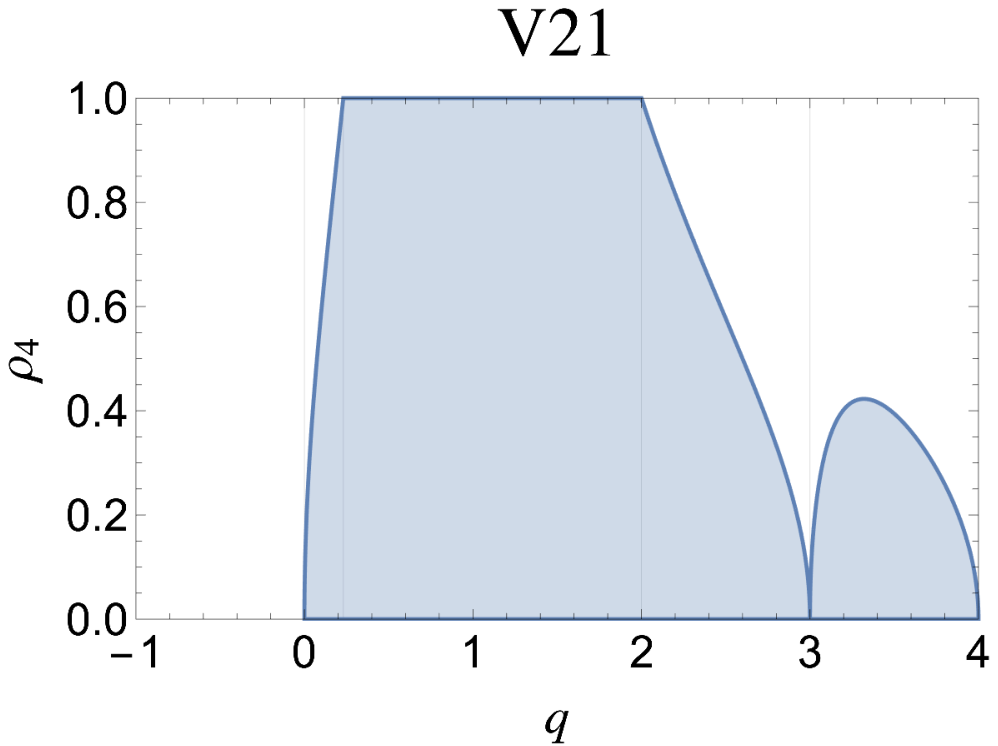}\hspace{25pt}
\includegraphics[width=0.45\textwidth]{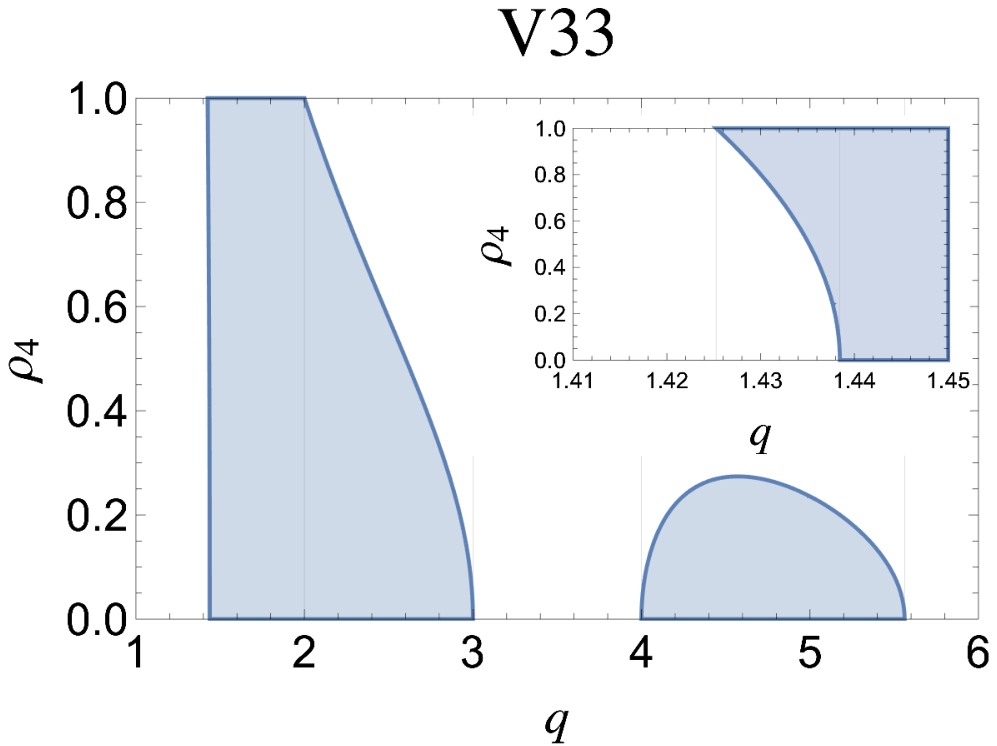}
\caption{Ranges of variation of $\rho_4$ for the $q=r$ solutions V1, V15, V21 and V33. \label{fig:Vsymq=r}}
\end{figure}

\vspace{.3cm}
Then we have the solutions V1, V15, V21 and V33 with $q=r$. They all possess a free parameter and correspond to lines of fixed points at fixed $q=r$ (see figure~\ref{fig:vsymq=rmod}). For V15, V21 and V33 the free parameter is $\rho_4$ itself (figure~\ref{fig:Vsymq=r}). The four solutions are

\noindent$\bullet$ V1 defined in the interval $q=r\in(2,3)$
\begin{equation}
\begin{aligned}
&\rho_{0,1}=\rho_{3,1}=\rho_{0,2}=\rho_{3,2}=0, \quad \rho_{1,1} = \rho_{1,2} = 1,   \quad \phi_1=-\phi_2\in [0,2\pi), \\
& \qquad \quad \rho_4=\sqrt{\frac{3-q}{q-1}}, \qquad \theta= (\pm)\frac{\pi}{2} ,\qquad \rho_6=(\pm)\sqrt{2\,\frac{q-2}{q-1}},
\end{aligned}
\end{equation}
$\bullet$ V15 defined in the interval $q=r \in (-1, 3)$
\begin{align}
\begin{split}
&\rho_{0,1}=\rho_{0,2}=0, \quad \rho_{3,1}=\rho_{3,2}=2x_1, \quad
 \theta=(\pm)\frac{\pi}{2}, \quad \rho_6=(\pm)\sqrt{1-\rho_4^2}\\
&x_1=x_2 = \pm\frac{1}{2}\sqrt{3-q+(q-1)\rho_4^2}, \quad y_1=-y_2=(\pm)\sqrt{1+q+(q-1)\rho_4^2} \\
& \hspace{80pt} 0 < \rho_4 < \begin{cases}
\sqrt{\frac{1+q}{1-q}},  &  -1 < q < 0, \\
1, &   \hspace{9pt} 0 \le q \le 2, \\
\sqrt{\frac{3-q}{q-1}}, & \hspace{9pt} 2 < q < 3,
\end{cases}
\end{split}
\end{align}
$\bullet$ V21 defined in the interval $q = r \in (0, 4)$
\begin{equation}
\begin{aligned}
&\rho_{0,1} = \rho_{0,2} = \pm \sqrt{1+\frac{q-1}{q-3}\rho_4^2}, \quad x_1 = x_2 = -\frac{q-4}{2} \rho_{0,1},  \quad  \theta=(\pm)\frac{\pi}{2}, \\
& y_1 = -y_2 = (\pm)\frac{1}{2}\sqrt{q(4-q)-\frac{(q-1)(q-2)^2}{q-3}\rho_4^2},\quad \rho_{3,1} = \rho_{3,2} = -(q-3)\rho_{0,1} ,\\
&\rho_6=(\pm)\sqrt{1-\rho_4^2},  \quad   0 < \rho_4 < \begin{cases}
\sqrt{\frac{q(q-3)(4-q)}{(q-1)(q-2)^2}}, & \hspace{5pt} 0 < q \le q^* \text{ and } 3<q<4, \\
1, & q^* < q \le 2, \\
\sqrt{\frac{3-q}{q-1}}, & \hspace{5pt} 2 < q < 3\,,
\end{cases}
\end{aligned}
\end{equation}
where $q^* \approx 0.231\ldots$ is the real root of the polynomial $q^3 - 6 q^2 + 10 q-2$. \\
\vspace{.3cm}
\noindent $\bullet$ V33 defined in the interval $q=r \in (q',3)\cup\left (4,\tfrac{7+\sqrt{17}}{2}\right )$ 
\begin{equation}
\begin{aligned}
& \rho_{0,1}=\rho_{0,2} = \rho_{3,1}=\rho_{3,2} = \pm\sqrt{\frac{q-3 +(q-1)\rho_4^2}{q^2 -5q+5}}, \qquad x_1 = x_2 = -\frac{q-4}{2}\rho_{0,1}, \\
& y_1 = -y_2 = (\pm)\frac{1}{2}\sqrt{\frac{(4-q) (q^2-7 q+8 )-(q-2)^2 (q-1) \rho_4^2}{q^2-5 q+5}}, \qquad \theta = (\pm) \frac{\pi}{2} , \\
& \rho_6 = (\pm)\sqrt{1-\rho_4^2}, \qquad \rho_4 \in \begin{cases}
\left ( \sqrt{\frac{(4-q) \left(q^2-7 q+8\right)}{(q-2)^2 (q-1)}}, 1 \right ), & q' < q < \frac{7-\sqrt{17}}{2},\\
\left (0, 1 \right ), & \frac{7-\sqrt{17}}{2} < q < 2, \\
\left (0, \sqrt{\frac{3-q}{q-1}} \right ), & 2 < q < 3, \\
\left (0, \sqrt{\frac{(4-q) \left(q^2-7 q+8\right)}{(q-2)^2 (q-1)}} \right ), & 4 < q < \frac{7+\sqrt{17}}{2},
\end{cases}
\end{aligned}
\end{equation}
where $q'\approx 1.425\ldots$ is the real root of the polynomial $q^3-8 q^2+22 q-18$. 

\vspace{.2cm}
Finally there are solutions defined only for isolated values of $q$ and $r$. The first is

\noindent$\bullet$ V13 defined for $r=q=2$
\begin{align}
\begin{split}
&\rho_{0,1}=\rho_{0,2}=\rho_{3,1}=\rho_{3,2}=x_1=x_2=\pm\sqrt{1-\rho_4^2} ,\quad \theta=(\pm)\frac{\pi}{2}  \\
&y_1=(\pm)1, \quad y_2=(\pm)1 , \quad \rho_6=(\pm)\sqrt{1-\rho_4^2} , \quad \rho_4\in(0,1)\,,
\end{split}
\end{align}
while the others correspond to less interesting (irrational) values of $q$ and $r$ and will not be listed.

\subsubsection{Solutions with $\rho_4=1$}
When $\rho_4=1$ (and then $\rho_6=0$), it can be seen that \eqref{uni9} allows to express $\theta$ in the form
\begin{equation}\label{Stheta}
\tan (\theta + \pi j) =  \frac{(q-2)\rho_{1,1}\cos \phi_1 + \rho_{3,1} + (r-2)\rho_{1,2}\cos \phi_2 + \rho_{3,2}}{(q-2) \rho_{1,1}\sin \phi_1 + (r-2)\rho_{1,2}\sin\phi_2} , \quad j = 0, 1\,,
\end{equation}
with the condition
\begin{equation}
\begin{aligned}
&[(q-2)\rho_{1,1}\cos \phi_1 + \rho_{3,1}]^2 + (q-2)^2 \rho_{1,1}^2\sin^2 \phi_1 \\
&\hspace{0.3\textwidth} = [(r-2)\rho_{1,2}\cos \phi_2 + \rho_{3,2}]^2 + (r-2)^2\rho_{1,2}^2\sin^2\phi_2 \,.
\end{aligned}
\end{equation}
If $(q-2)\rho_{1,1}\cos \phi_1 + \rho_{3,1}=
(q-2) \rho_{1,1}\sin \phi_1 =
(r-2)\rho_{1,2}\cos \phi_2 + \rho_{3,2} =
(r-2)\rho_{1,2}\sin\phi_2 =0$, $\theta$ becomes a free parameter. We can solve for the remaining parameters, which satisfy the equations
\begin{align}
\begin{aligned}
0&=(q-4)\rho_{0,1}^2+2\rho_{1,1}\rho_{0,1}\cos\phi_1, & \quad 0&=(r-4)\rho_{0,2}^2+2\rho_{1,2}\rho_{0,2}\cos\phi_2, \\
1&=(q-3)\rho_{0,1}^2+\rho_{1,1}^2, &\quad 1&=(r-3)\rho_{0,2}^2+\rho_{1,2}^2, \\
1-r&=(q-3)\rho_{1,1}^2+2\rho_{1,1}\rho_{3,1}\cos\phi_1, & \quad 1-q&=(r-3)\rho_{1,2}^2+2\rho_{1,2}\rho_{3,2}\cos\phi_2, \\
2-r&=(q-2)\rho_{1,1}^2+\rho_{3,1}^2, &\quad 2-q &=(r-2)\rho_{1,2}^2+\rho_{3,2}^2\,.
\end{aligned}
\end{align}
Notice that the last two equations imply that no solution in this class exists if both $q$ and $r$ are larger than 2.

\begin{figure}
\centering
\includegraphics[width=0.32\textwidth]{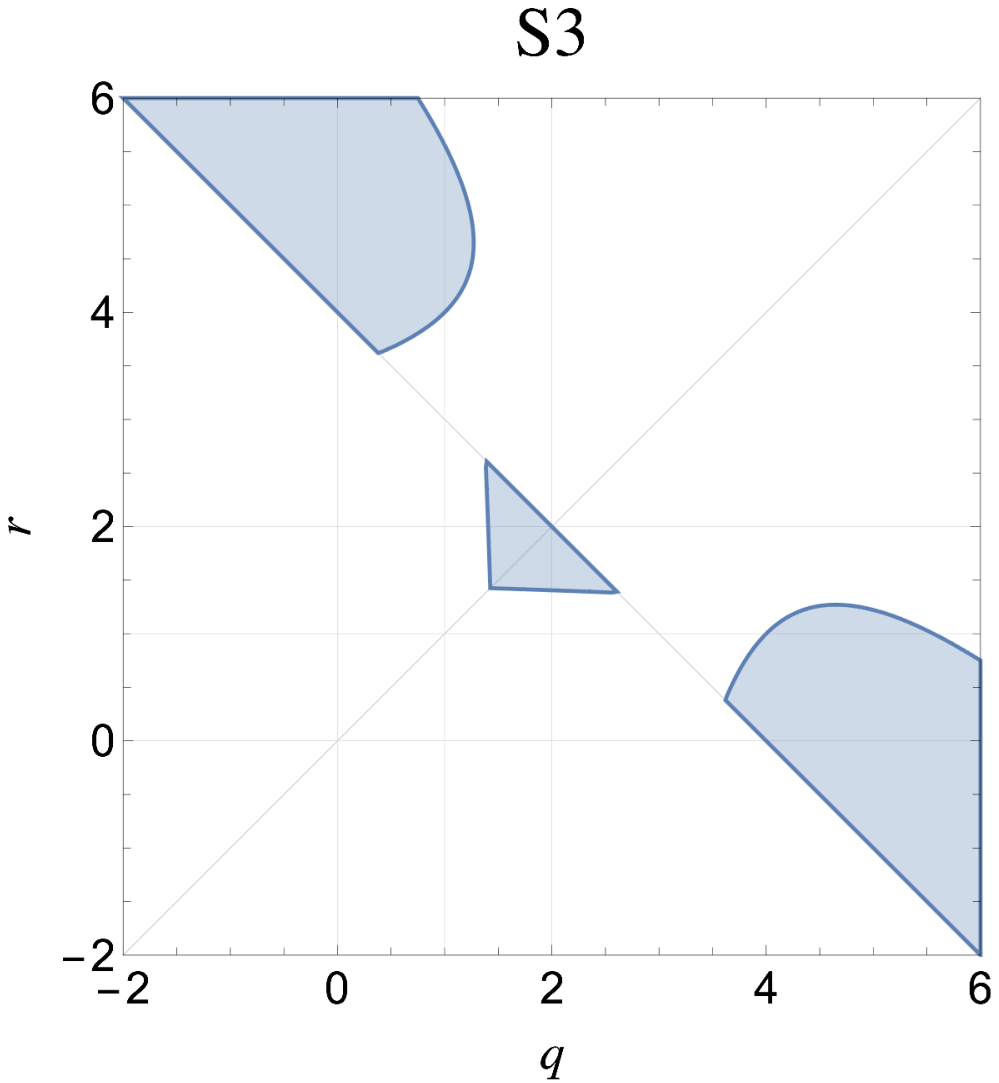} \,  
\includegraphics[width=0.32\textwidth]{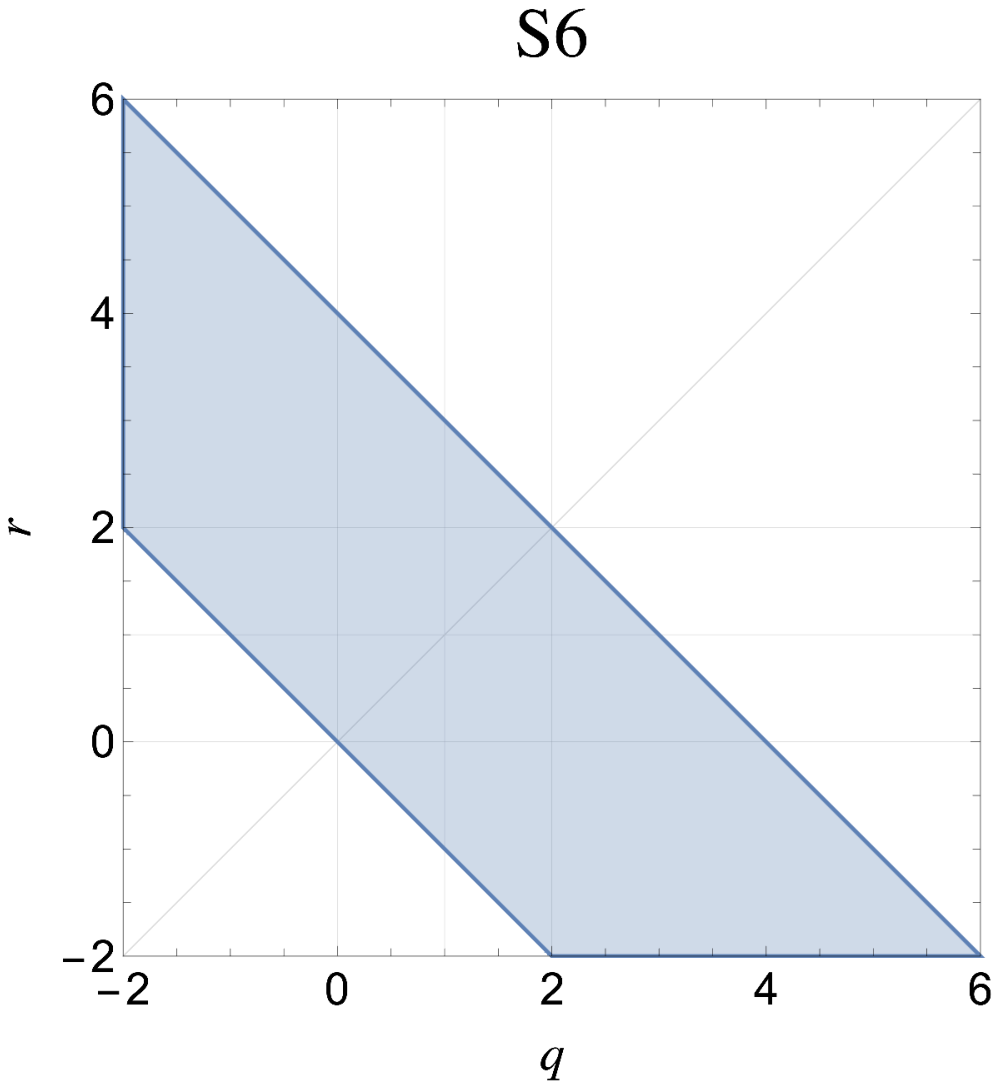} \, 
\includegraphics[width=0.32\textwidth]{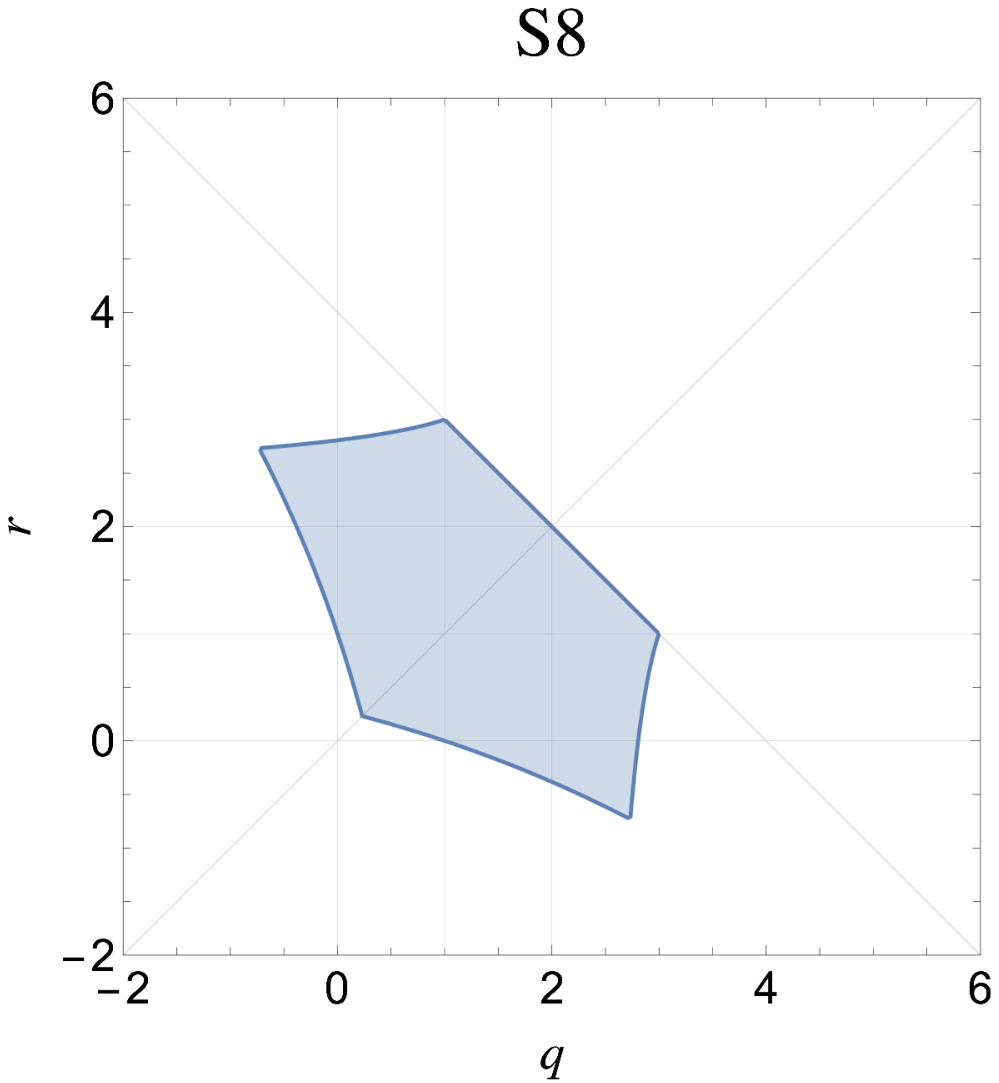}
\caption{Domains of definition in the $q$-$r$ plane for the solutions S3, S6, and S8\label{fig:Ssymgen}}
\end{figure}
We list here the exchange-invariant solutions, starting with

\noindent $\bullet$ S1 defined for $q+r=4$
\begin{align}
\begin{split}
\rho_{0,1}=\rho_{0,2}=\rho_{3,1}=\rho_{3,2}=0  \, \, , \, \, \phi_1,\phi_2 \in[0,2\pi) \, \, , \, \, \rho_{1,1}=\rho_{1,2}=1\,.
\end{split}
\end{align}
The three solutions S3, S6 and S8 are instead defined for the values of $q$ and $r$ shown in figure \ref{fig:Ssymgen}. They read

\noindent$\bullet$ S3 
\begin{align}
\begin{split}
&\rho_{0,1}=\rho_{3,1}=(\pm)\sqrt{ \frac{q+r-4}{q^2-5q+5}}, \qquad \rho_{0,2}=\rho_{3,2}=(\pm)\sqrt{ \frac{q+r-4}{r^2-5r+5}},\\
&x_1=- \frac{q-4}{2}\rho_{0,1}, \quad y_1=(\pm)\frac{1}{2} \sqrt{\frac{(4-q) \left(q^2-7 q+8\right)-(q-2)^2 (r-1)}{q^2-5 q+5}},\\
&x_2=-\frac{r-4}{2}\rho_{0,2}, \quad y_2=(\pm)\frac{1}{2} \sqrt{\frac{(4-r) \left(r^2-7 r+8\right)-(r-2)^2(q-1)}{r^2-5 r+5}},
\end{split}
\end{align}
$\bullet$ S6
\begin{align}
\begin{split}
&\rho_{0,1}=\rho_{0,2}=0 , \quad \rho_{3,1}=2x_1 \, \, , \, \, \rho_{3,2}=2x_2\\
&x_1=(\pm) \frac{1}{2}\sqrt{4-q-r} \, \, , \, \, x_2=(\pm)\frac{1}{2}\sqrt{4-q-r}\\
&y_1=(\pm)\frac{1}{2}\sqrt{q+r} \, \, , \, \, y_2=(\pm)\frac{1}{2}\sqrt{q+r},
\end{split}
\end{align}
$\bullet$ S8
\begin{align}
\begin{split}
&\rho_{0,1}=(\pm)\sqrt{\frac{q+r-4}{q-3}}  \quad \rho_{0,2}=(\pm)\sqrt{\frac{q+r-4}{r-3}} \\
&x_1= -\frac{q-4}{2}\rho_{0,1} , \quad \rho_{3,1}=-(q-3)\rho_{0,1} , \quad x_2=-\frac{r-4}{2}\rho_{0,2} , \quad \rho_{3,2}=-(r-3)\rho_{0,2}\\
&y_1=(\pm)\frac{1}{2}\sqrt{\frac{4-r(q-2)^2-q(q-4)^2}{q-3}}, \quad y_2=(\pm)\frac{1}{2}\sqrt{\frac{4-q(r-2)^2-r(r-4)^2}{r-3}}.
\end{split}
\end{align}

Finally, we have solutions defined for isolated values of $q$ and $r$. The first one is

\noindent$\bullet$ S9 defined for $q=r=2$
\begin{align}
\begin{split}
\rho_{0,1}=\rho_{0,2}=\rho_{3,1}=\rho_{3,2}=0  \, \, , \, \, \phi_1,\phi_2, \theta \in[0,2\pi) \, \, , \, \, \rho_{1,1}=\rho_{1,2}=1\,,
\end{split}
\end{align}
while the others are defined for less interesting values of $q$ and $r$ (irrational or zero) and will not be listed.

\subsection{Some implications}
We notice first of all that for $q=r=2$ the Hamiltonian (\ref{H_qr}) becomes that of the Ashkin-Teller model (two coupled Ising models) \cite{AT}. The Ashkin-Teller model allows for lines of fixed points -- along which the central charge is $c=1$ and the critical exponents vary continuously -- identified perturbatively in \cite{JKKN,Kadanoff} and exactly in \cite{ising_vector}. Also in our present formulation we see that at $q=r=2$ the solutions V15, V21, V33, V13, S1 and S9 all possess a free parameter providing the coordinate along a critical line. 

Excluding the Ashkin-Teller case, we also see that the only critical points corresponding to coupled models with $q$ and $r$ integers larger than 1 are provided by solution V5 for $(q,r)=(2,5)$ and solution V33 for $q=r=5$. Both V5 and V33 appear to be related to solution V of table~\ref{qsol} for a single Potts model. One implication is that for criticality in coupled ferromagnets we are left with Ashkin-Teller only. As a particular case, these results confirm the conclusion of theoretical and numerical studies \cite{Vaysburd,Pujol,DJLP,GC} which found no fixed points in two coupled Potts ferromagnets with $q=r=3,4,\ldots$ .

\begin{figure}[t]
\centering
\includegraphics[width=0.7\textwidth]{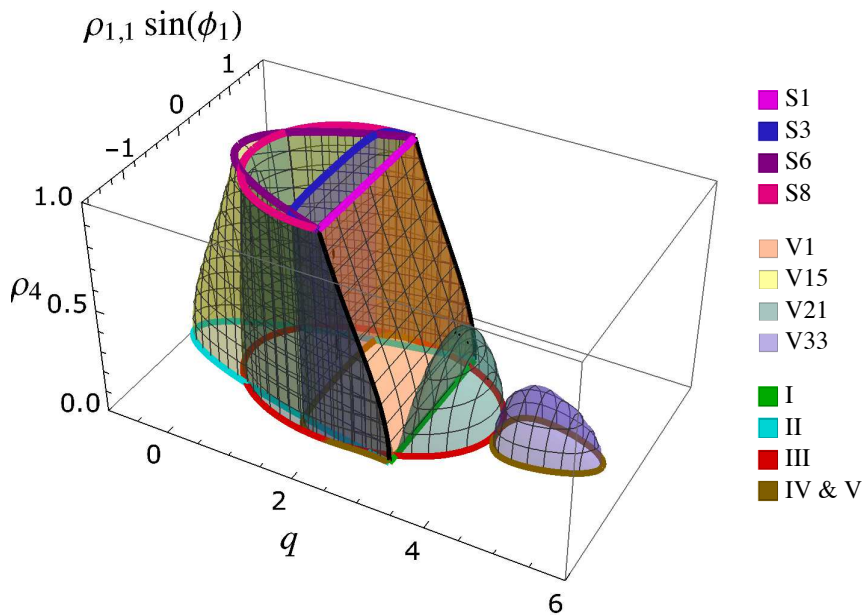}
\caption{The solutions V1, V15, V21, and V33 in the parameter subspace ($\rho_{1,1}\sin\phi_1$, $\rho_4$) along with the decoupled solutions I, II, III, IV and V, and the strongly coupled solution S1, S3, S6, S8 for $q=r$. All the surfaces corresponding to the V-type solutions meet along the black lines for $2<q<3$. \label{fig:vsymq=rmod}}
\end{figure}

The nontrivial phenomenon of a critical line with continuously varying exponents for fixed symmetry provided by the Ashkin-Teller model is made possible by the fact that at criticality this model renormalizes on the Gaussian model (free scalar boson), which in two dimensions allows for a continuous spectrum of scaling dimensions \cite{dFMS}. The same free boson accounts for the line of fixed points of a single $3$-state Potts model (solution I of table~\ref{qsol}) \cite{DT1}, and is expected to generate all lines of fixed points at fixed symmetry in two dimensions. In particular, it should be contained in the field theories corresponding to the $q=r$ solutions V1, V15, V21 and V33, which span surfaces of fixed points as $q$ varies (figure~\ref{fig:vsymq=rmod}), and then lines of fixed points for $q$ fixed. Quite interestingly a solvable square lattice realization of the $q=r$ model originally obtained in \cite{Perk} and further studied in \cite{MN,FJ} has been found in \cite{VJS} to possess a continuous spectrum of critical exponents associated to a free boson. This lattice solution terminates at $q=4$ and has $q=3$ as a decoupling point; it should then be related to our solution V21. It can be seen also from figure~\ref{fig:vsymq=rmod} that in the decoupling limit $\rho_4\to 0$ the solution V21 gives solution III for the two $q$-state models, and we saw in section~\ref{single} that solution III can describe both the ferromagnet and the square lattice antiferromagnet (recall table~\ref{summary}). In \cite{VJS} the lattice solution results from the coupling of  two antiferromagnets; on the other hand the central charge proposed in \cite{FJ,VJS} gives at the decoupling point $q=3$ the value $8/5$, which is that of two ferromagnets. This point will deserve further investigation.

It was shown in \cite{DR,DJNS} that the requirement of lattice self-duality for the Hamiltonian (\ref{H_qr}) is satisfied on a one-dimensional manifold in the coupling space of the lattice model. This requirement is distinct from that of scale invariance, so that the room for comparison with our results is necessarily limited. We can nonetheless observe that our results show that scale invariance occurs in the field theoretical coupling space at most on one-dimensional manifolds, which correspond to the lines of fixed points at fixed $q=r$ discussed in this subsection.

\section{Correlated percolation}
\label{correlated}
Correlated percolation in the $q$-state Potts ferromagnet was studied in \cite{CP} considering the coupling to an $r$-state Potts model realized by the Hamiltonian (\ref{H_qr}) with $J_2=0$. The site variables of the $r$-component are auxiliary and are eventually eliminated by the limit $r\to 1$. They allow a generalization of the FK expansion in which the clusters are made of connected bonds placed with probability (\ref{prob}) between nearest neighbors with the same value of the $q$-state Potts site variable. This comes from the fact that in (\ref{H_qr}) $J$  is the coupling of the auxiliary variables to the $q$-state Potts variables. The approximated RG analysis of \cite{CP} gave two fixed points for the coupled ($J\neq 0$) model for $r\to 1$. Both of them have $J_1$ equal to the critical value $J_c$ of the $q$-state Potts ferromagnet, consistently with the expectation for a fixed point of the $r\to 1$ model. One of the fixed points, for $J=J_c$, was argued to be repulsive and to correspond to the FK clusters. The other fixed point, for $J=J^*>J_c$, was argued to be attractive and to rule the critical behavior of spin clusters, which by definition correspond to $J=\infty$ ($p=1$). The lines spanned by the two fixed points as a function of $q$ were conjectured in \cite{CP} to coalesce and terminate at $q=4$, although this could not actually be seen within the approximated RG method. 

We are now in the position of discussing Potts correlated percolation in light of our exact fixed point solutions. Observe first of all that for $r=1$ the equations (\ref{uni1}), (\ref{uni3}), (\ref{uni5}) and (\ref{uni7}) exactly reproduce the system (\ref{uniq1})-(\ref{uniq4}) which determines the fixed points of a decoupled $q$-state Potts model. This implies that, even in the coupled model ($\rho_4\neq 0$), for $r\to 1$ the amplitudes of the $q$-state sector -- namely the amplitudes $S_{i,1}$, $i=0,1,2,3$, involving only excitations in the $q$-state sector -- coincide with one of the solutions of table~\ref{qsol} for the decoupled $q$-state Potts model. This can indeed be checked plugging $r=1$ in the solutions of section~\ref{fpsolutions}. On the other hand, it must be taken into account that, even for $r\to 1$, the remaining amplitudes -- involving excitations in the auxiliary $r$-state sector -- cannot be forgotten, since also the auxiliary degrees of freedom will normally contribute to the results\footnote{This is immediately understood thinking to the basic example of random percolation as the limit $r\to 1$ of the $r$-state Potts model. In this case all the degrees of freedom are auxiliary, but nevertheless determine all the percolative properties. Extensive illustrations in the scattering formalism for the off-critical case are given in \cite{DC98,isingperc,DVC,DV_droplets,DV_crossing}.}. It follows that, taking into account the full set of amplitudes that determine a fixed point solution, the range of $q$ in which the solution is defined for $r\to 1$ is in general smaller than that given in table~\ref{qsol} for the amplitudes of the $q$-state sector. This is due to the fact that in general the $r$-state (auxiliary) degrees of freedom are not critical in the full range in which the $q$-state degrees of freedom are critical. In particular, inspection of the solutions of section~\ref{fpsolutions} and appendix~\ref{noninv} shows that in the $r\to 1$ coupled model there is no critical line continuously defined in the whole range $q\in[2,4]$. This excludes, in particular, that spin clusters can be fully described through some analytic continuation performed in this range, consistently with the numerical findings of \cite{DPSV,KEWI}. The possibility remains, however, that analytic continuation holds for some specific quantity that can be evaluated directly at $r=1$, where the number of auxiliary degrees of freedom is strictly zero\footnote{We further illustrate this point in appendix~\ref{eigen}.} and one can recover the branches of solution III of table~\ref{qsol} in their full range of definition $q\in[0,4]$.  

In light of our present results, it seems interesting to recall that different convergenge patterns of spin clusters towards a fixed point were observed below and above $q\approx 2.5$ in the Monte Carlo simulations performed in \cite{DPSV}, where the range $q\in[1,4]$ was scanned in steps\footnote{Noninteger values of $q$ have been studied numerically also with transfer matrix techniques in \cite{DJS1,DJS2}, where the data have been used to conjecture domain wall exponents (see also \cite{VJ}).} of 0.25. This can be compared with our finding that in the $r\to 1$ coupled model there is a discontinuity in the critical properties at $q=3$, since solution S8 provides a critical line in the range $q\in[0,3]$, and solution $\tilde{\textrm{S}}$5 provides a critical line in the range $q\in(3,4]$; here the tilde indicates the solution obtained from (\ref{S5}) under the exchanges (\ref{exchange}). As $r\to 1$, both S8 and $\tilde{\textrm{S}}$5 tend to solution III of table~\ref{qsol} in the $q$-state sector and in their respective domains of definition.

\section{Conclusion}
In this paper we used scale invariant scattering theory to exactly determine the RG fixed points of a $q$-state Potts model coupled to an $r$-state Potts model in two dimensions. The analysis is performed directy in the continuum limit, looking for conformally invariant points within the basis of particle excitations. The results are obtained for continuous values of $q$ and $r$, and include both ferromagnetic and antiferromagnetic interactions. For $q$ and $r$ integers larger than 1 the room for criticality turns out to be quite limited, and for coupled ferromagnets one is only left with the Ashkin-Teller case ($q=r=2$), which allows for lines of fixed points. Once the number of states is allowed to be noninteger, an interesting phenomenon is the appearance of surfaces of fixed points spanned as $q=r$ varies. One of these surfaces appears to correspond to the case of coupled square lattice antiferromagnets previously studied in the lattice literature \cite{Perk,MN,FJ,VJS}. 

As a particular case, we have exact access to the limit $r\to 1$, which is related to Potts correlated percolation and up to now could only be studied through the approximated RG calculations of \cite{CP}. We find that in the coupled model there is no critical line along which the degrees of freedom of the $q$-state sector and those of the $(r=1+\epsilon)$-state sector are critical with continuity in the whole range $q\in[2,4]$. This implies that the conjecture that Potts spin clusters can be described as an analytic continuation of FK clusters cannot hold for all cluster properties, and explains why the conjecture was numerically found to fail for the quantities considered in \cite{DPSV,KEWI}. On the other hand, the conjecture could hold for some specific property that can be determined directly at $r=1$, where the number of degrees of freedom in the $r$-state sector is strictly zero and the constraints coming from that sector can be ignored. This might still allow the formula for the fractal dimension of spin clusters, conjectured in \cite{Vanderzande} and up to now consistent with numerical simulations, to be exact.

\appendix
\section{Exchange-noninvariant solutions}
\label{noninv}
We list in this appendix the solutions of the fixed point equations (\ref{uni1})-(\ref{uni11}) which are noninvariant under the exchange operation (\ref{exchange}). This operation maps each of the solutions below into another solution, which we do not list.

\begin{figure}[t]
\centering
\includegraphics[width=0.32\textwidth]{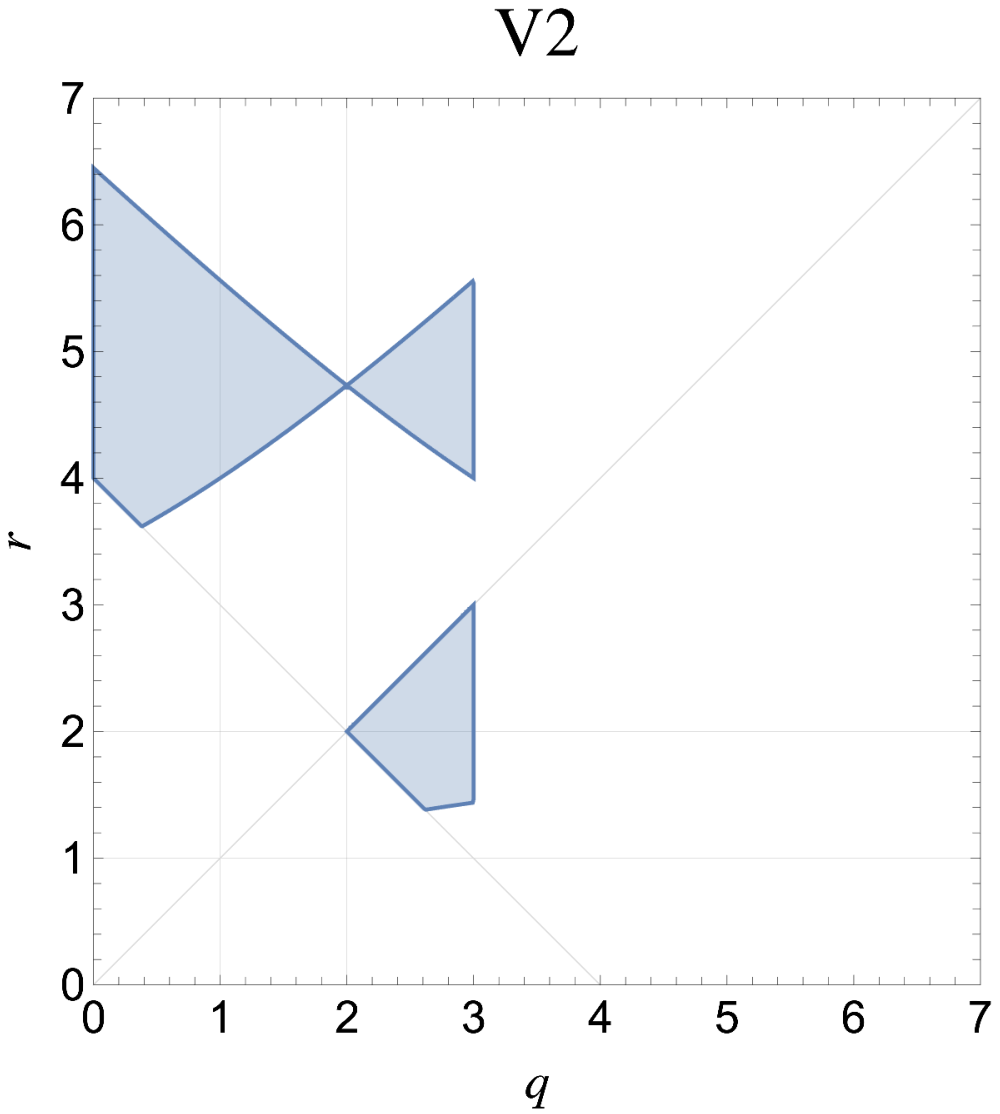}\,
\includegraphics[width=0.32\textwidth]{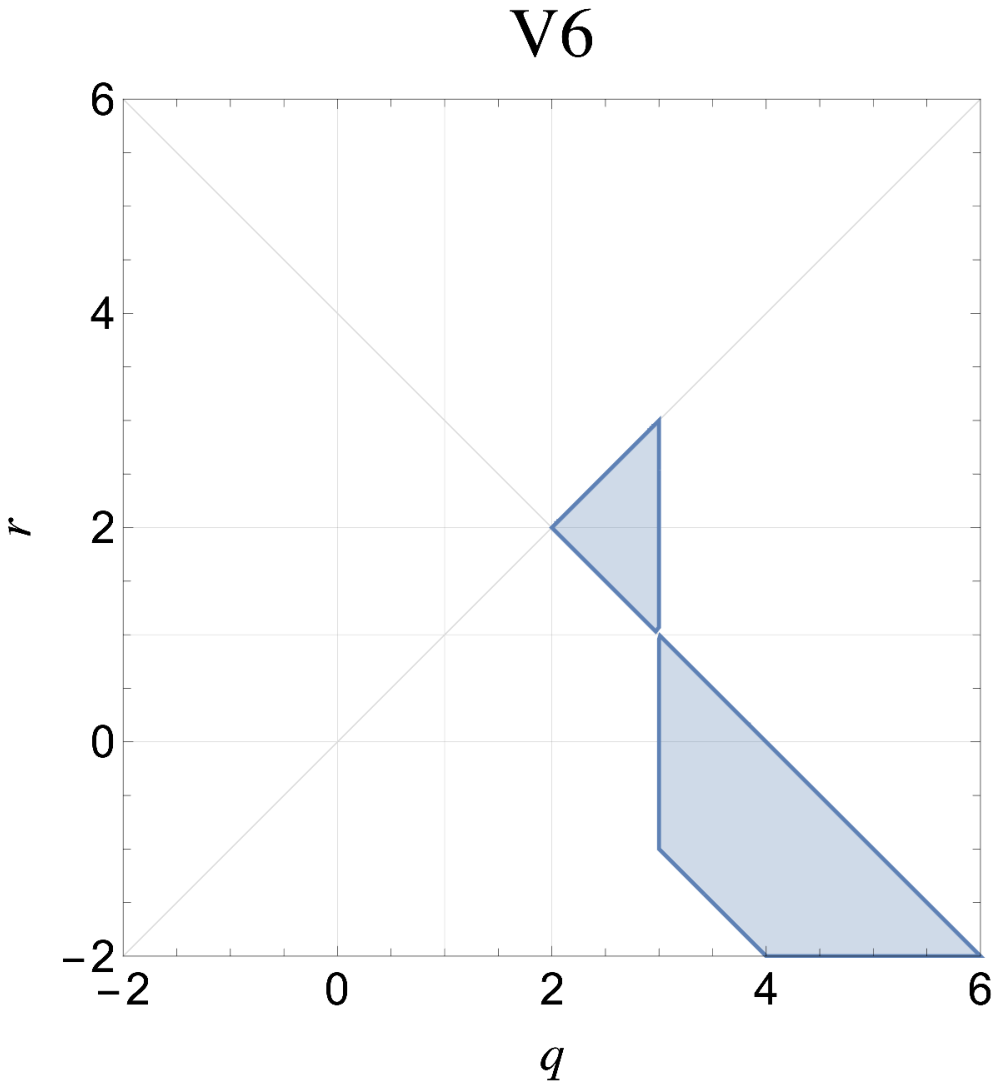}\, 
\includegraphics[width=0.32\textwidth]{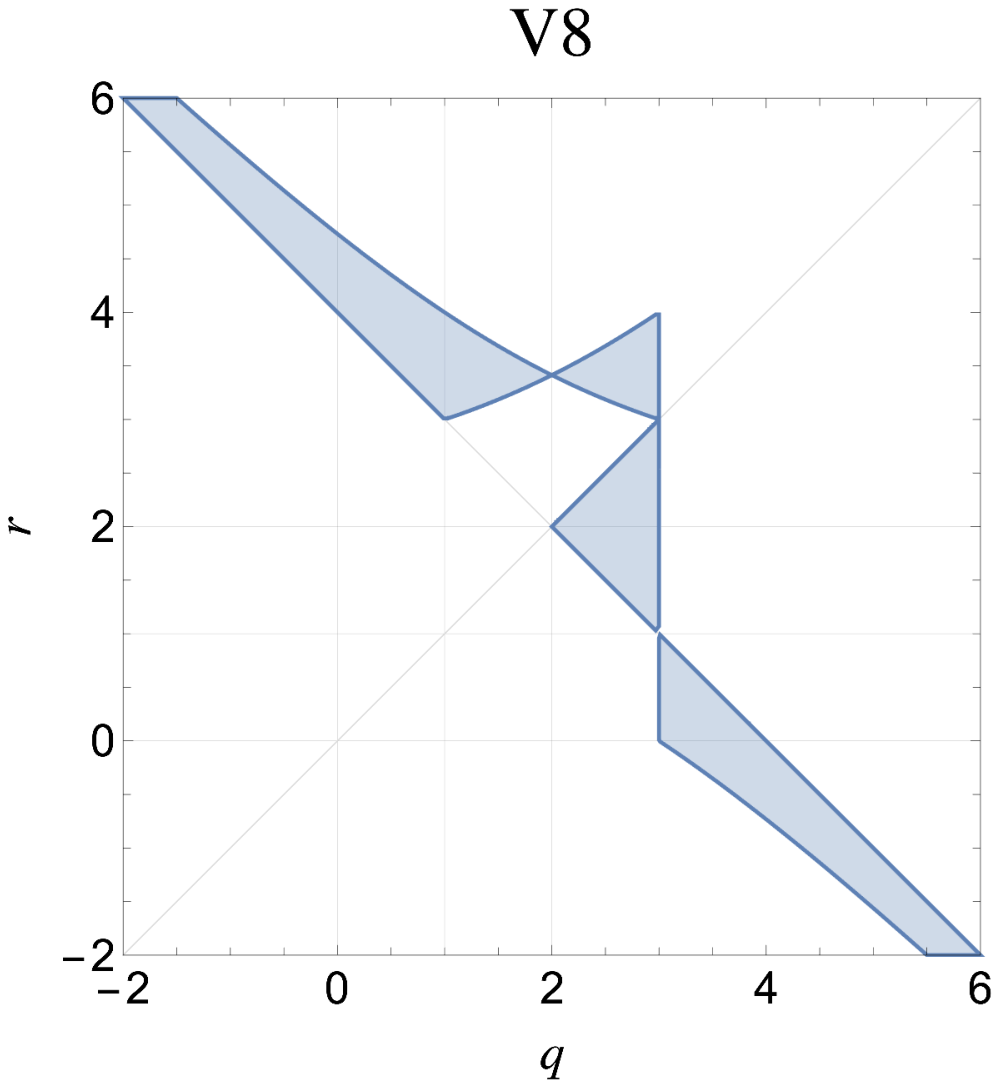}\\[3ex]
\includegraphics[width=0.32\textwidth]{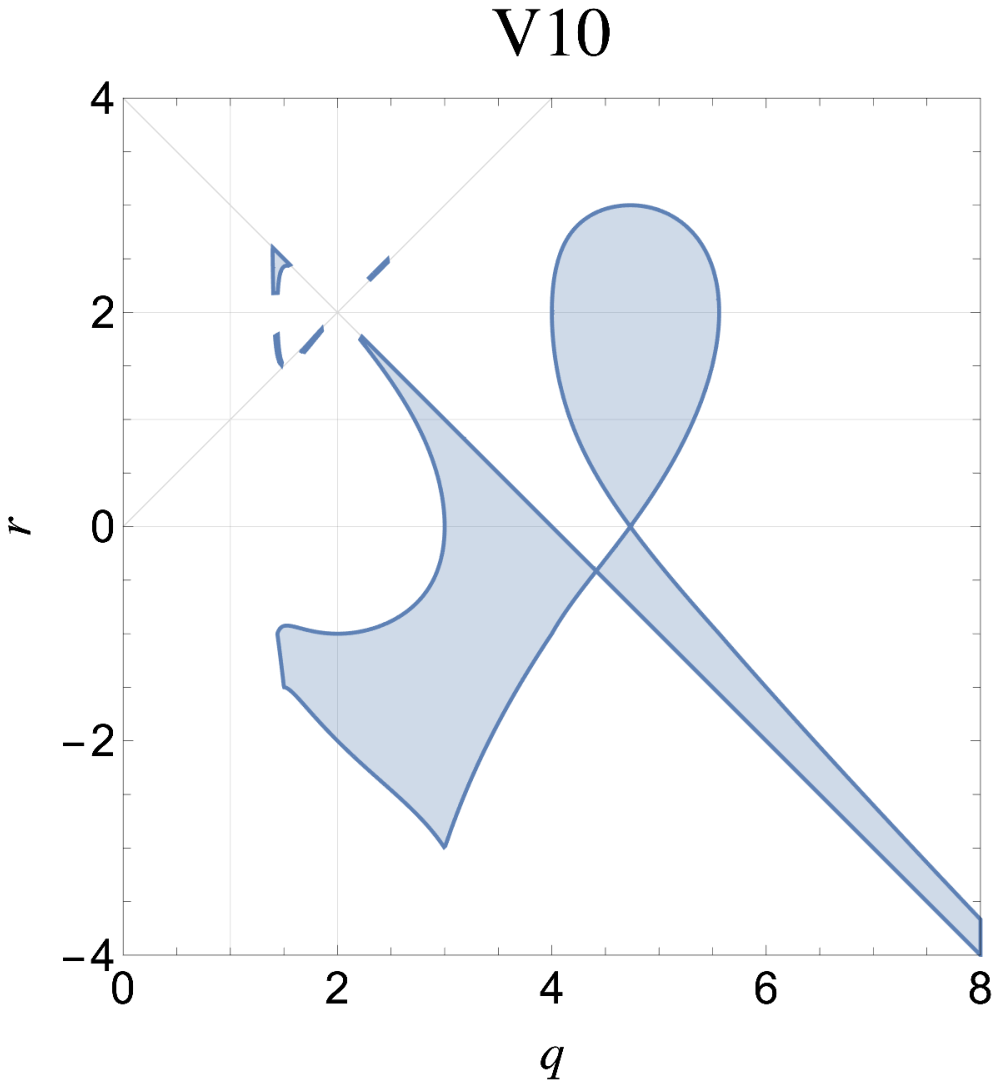}\,
\includegraphics[width=0.32\textwidth]{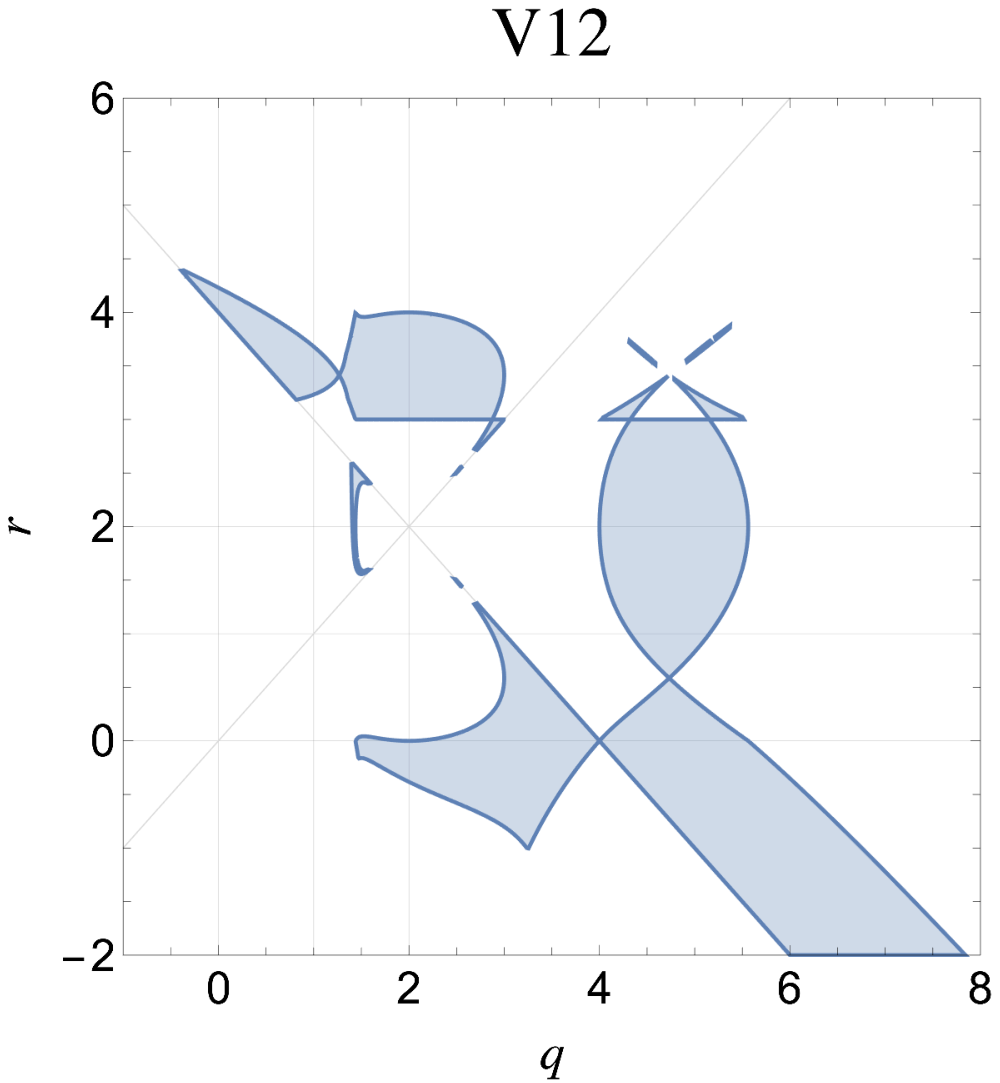}\,
\includegraphics[width=0.32\textwidth]{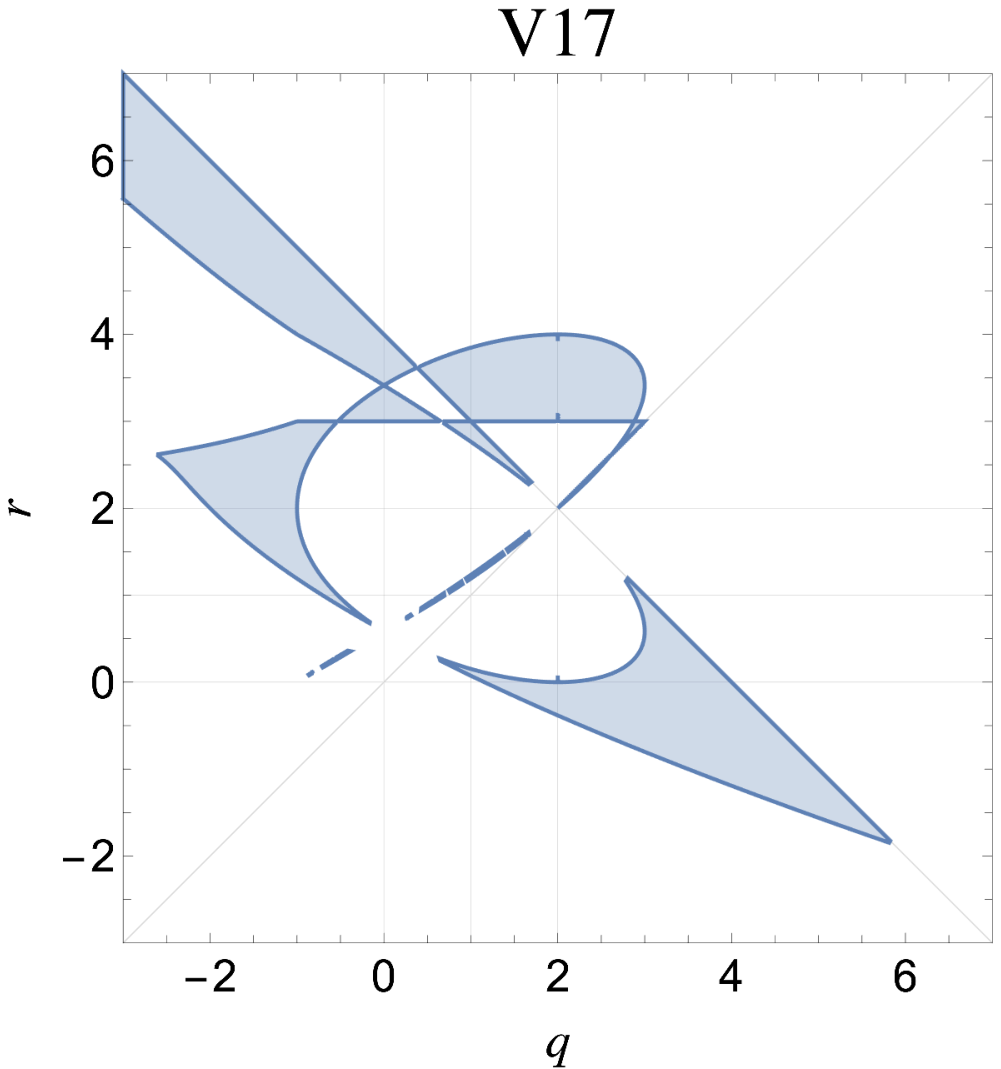}
\caption{Domains of definition of the solutions V2, V6, V8, V10, V12 and V17.\label{fig:Vasymgen}}
\end{figure}

We first have the following solutions with varying $\rho_4$ and domains of definition shown in figure \ref{fig:Vasymgen}:

\noindent$\bullet$ V2
\begin{equation}
\begin{aligned}
&\rho_{0,1}=\rho_{3,1}=0 , \quad \rho_{0,2}=\rho_{3,2} = \pm \sqrt{\frac{(r-q)(r+q-4)}{(r-1)(r^2 -5r +5)}}, \\
&x_1=-\frac{(r^2-6 r+6)}{2 (q-2)}\rho_{0,2} , \quad x_2 = -\frac{r-4}{2}\rho_{0,2} ,\quad \theta=(\pm)\frac{\pi}{2},\\
&y_1=-\frac{r-2}{q-2}y_2 , \quad \rho_6=(\pm)\sqrt{\frac{q+r-4}{r-1}} \, \, , \, \, \rho_4=\sqrt{\frac{3-q}{r-1}},\\
&y_2=(\pm) \frac{1}{2} \sqrt{\frac{(q-3) (q-1)}{r-1}+\frac{(q-4) q-r+5}{r^2-5r+5}-r+6},
\end{aligned}
\end{equation}
$\bullet$ V6
\begin{align}
\begin{split}
&\rho_{0,1}=\rho_{0,2}=\rho_{3,1}=0  , \qquad \rho_{3,2}=2x_2 , \qquad \theta=(\pm)\frac{\pi}{2},\\
&x_1=\frac{r}{q-2}x_2 , \qquad x_2=\pm \frac{1}{2}\sqrt{\frac{(q-r)(q+r-4)}{r-1}} , \qquad \rho_4=\sqrt{\frac{3-q}{r-1}},\\
&y_1=-\frac{r-2}{q-2}y_2 , \qquad y_2=(\pm) \frac{1}{2}\sqrt{\frac{r^2-(q-2)^2}{r-1}} , \qquad \rho_6=(\pm)\sqrt{\frac{r+q-4}{r-1}},
\end{split}
\end{align}
$\bullet$ V8
\begin{equation}
\begin{aligned}
&\rho_{0,1}=\rho_{3,1}=0, \quad\rho_{0,2}=\pm \sqrt{\frac{(r-q)(q+r-4)}{(r-1)(r-3)}} , \quad \rho_{3,2}=-(r-3)\rho_{0,2} ,\quad \theta=(\pm)\frac{\pi}{2} ,\\
& x_1=-\frac{r^2-4r+2}{2(q-2)}\rho_{0,2}, \quad x_2 = -\frac{r-4}{2}\rho_{0,2}, \quad \rho_4=\sqrt{\frac{3-q}{r-1}}, \quad \rho_6=(\pm)\sqrt{\frac{q+r-4}{r-1}},\\
&y_1=-\frac{r-2}{q-2}y_2, \quad y_2=(\pm)\frac{1}{2}\sqrt{\frac{(q-3) (q-1)}{(r-3) (r-1)}(r-2)^2 -r(r-4)},
\end{aligned} 
\end{equation}

$\bullet$ V10
\begin{align}
\begin{split}
&\rho_{0,1}=\rho_{3,1}=  \pm r \sqrt{\frac{(q-r) (q+r-4)}{(q-1) (q^2-5 q+5) r^2+(q^2-6 q+6)^2 (r-1)}}, \quad \rho_{0,2}=0,\\
&x_1=-\frac{q-4}{2} \rho_{0,1}, \quad x_2=-\frac{q^2-6q+6}{2r} \rho_{0,1}, \quad  \rho_{3,2}=2x_2, \quad \theta=(\pm)\frac{\pi}{2},\\
&y_1=(\pm)\frac{r-2}{2} \sqrt{\frac{(q-2)^2 r^2-(q^2-6 q+6)^2}{(q-1) (q^2-5 q+5) r^2+(r-1)(q^2-6 q+6)^2}}, \quad y_2=-\frac{q-2}{r-2}y_1,\\
&\rho_4=\sqrt{\frac{r^2(3-r)(q^2-5 q+5) + (3-q) (q^2-6 q+6)^2}{r^2(q-1)(q^2-5 q+5) + (r-1)(q^2-6q+6)^2}}, \quad \rho_6=(\pm)\sqrt{1-\rho_4^2},
\end{split}
\end{align}
$\bullet$ V12
\begin{align}
\begin{split}
&\rho_{0,1}=\rho_{3,1}= \pm (r^2-4r+2) \alpha(q,r), \quad \rho_{0,2}= \pm (q^2 -6q +6)\alpha(q,r), \quad \theta=(\pm)\frac{\pi}{2},\\
&x_1= -\frac{q-4}{2}\rho_{0,1}, \quad x_2= -\frac{(r-4)}{2}\rho_{0,2} , \quad \rho_{3,2}= -(r-3) \rho_{0,2}, \quad \rho_6 = \pm \sqrt{1- \rho_4^2},\\
&y_1=(\pm) \frac{r-2}{2}\beta(q,r), \quad y_2= -\frac{q-2}{r-2}y_1, \quad \rho_4 = \sqrt{\frac{(r^2-4 r+2)^2-\frac{(q-3) (q^2-6 q+6)^2}{q^2-5 q+5}}{\frac{(q^2-6 q+6)^2 (r-1)}{q^2-5q+5}-\frac{(q-1) (r^2-4 r+2)^2}{r-3}}},\\
&\alpha(q,r)=\sqrt{\frac{(r-q) (q+r-4)}{(q^2-6 q+6)^2 (r-3) (r-1)-(q-1) (q^2-5 q+5) (r^2-4 r+2)^2}},\\
&\beta(q,r)=\sqrt{\frac{(q^2-6 q+6)^2 (r-2)^2-(q-2)^2 (r^2-4 r+2)^2}{(q^2-6 q+6)^2 (r-3) (r-1)-(q-1) (q^2-5q+5) (r^2-4 r+2)^2}},
\end{split}
\end{align}
$\bullet$ V17
\begin{align}
\begin{split}
&\rho_{0,1}=0 , \quad \rho_{0,2}=q \sqrt{\frac{ (q-r)(4-q-r)}{q^2 (r-3) (r-1)+(q-1) (r^2-4 r+2)^2}} , \quad \theta=(\pm)\frac{\pi}{2} , \\
&x_1=-\frac{r^2-	4r+2}{2q} \rho_{0,2} , \quad
x_2=-\frac{r-4}{2}\rho_{0,2} ,  \quad \rho_{3,1}=2x_1 , \quad \rho_{3,2}=-(r-3)\rho_{0,2} ,\\
&y_1=(\pm)\frac{r-2}{2} \sqrt{\frac{q^2 (r-2)^2-(r^2-4 r+2)^2}{q^2 (r-3) (r-1)+(q-1) (r^2-4 r+2)^2}}, \quad y_2 = -\frac{q-2}{r-2}y_1,\\
&\rho_4=\sqrt{\frac{(r-3) \left[(3-q) q^2-(r^2-4 r+2)^2\right]}{q^2 (r-3) (r-1)+(q-1) (r^2-4 r+2)^2}}, \quad \rho_6 = (\pm) \sqrt{1- \rho_4^2}.
\end{split}
\end{align}

\begin{figure}
\centering
\includegraphics[width=0.32\textwidth]{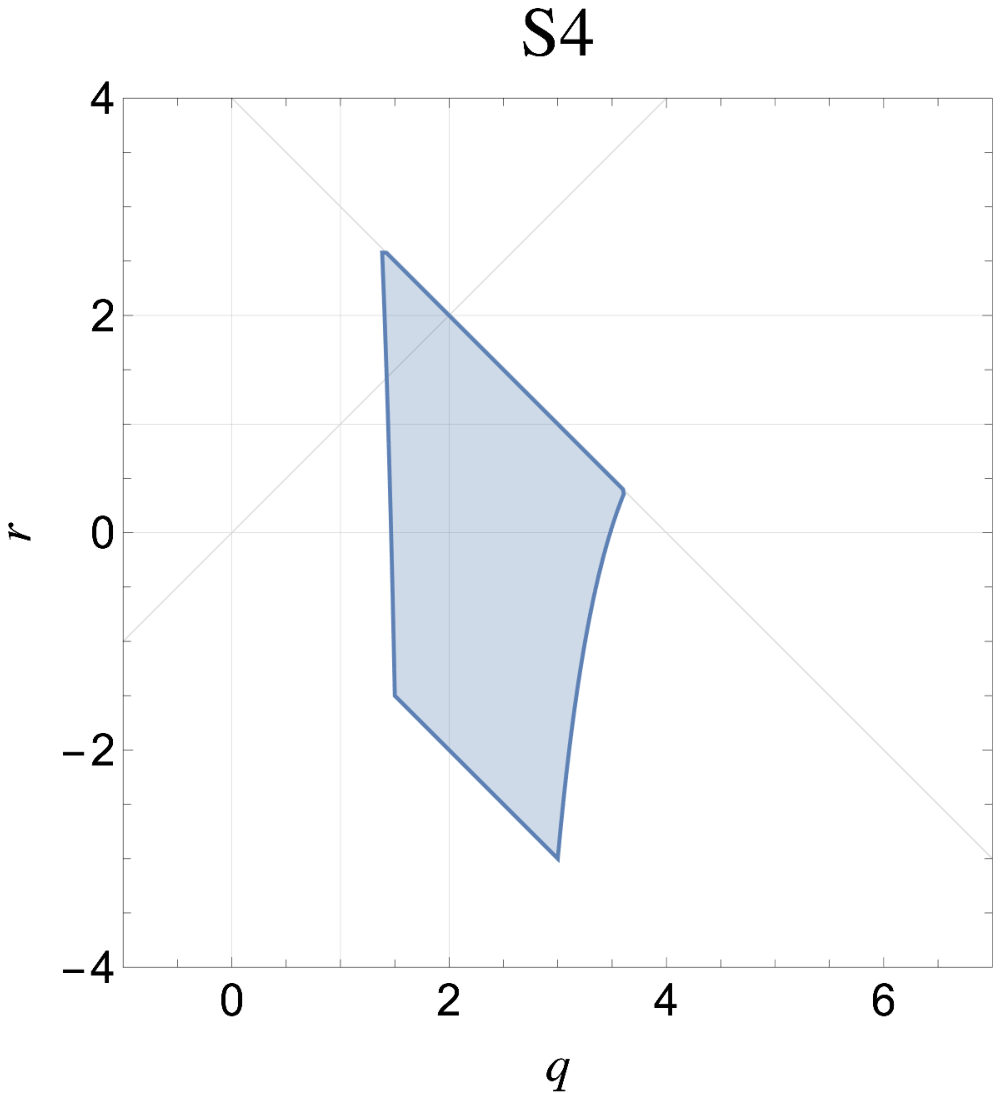} \,  
\includegraphics[width=0.32\textwidth]{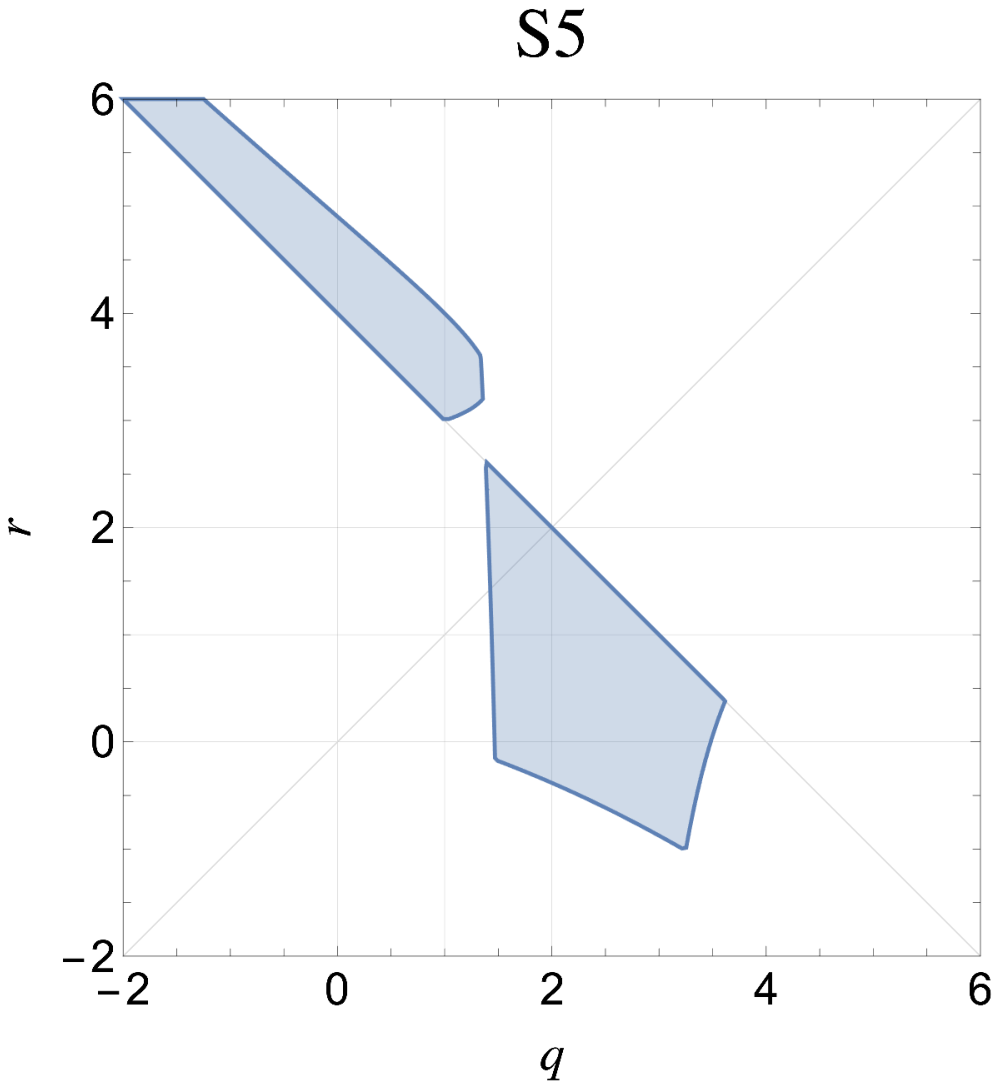} \, 
\includegraphics[width=0.32\textwidth]{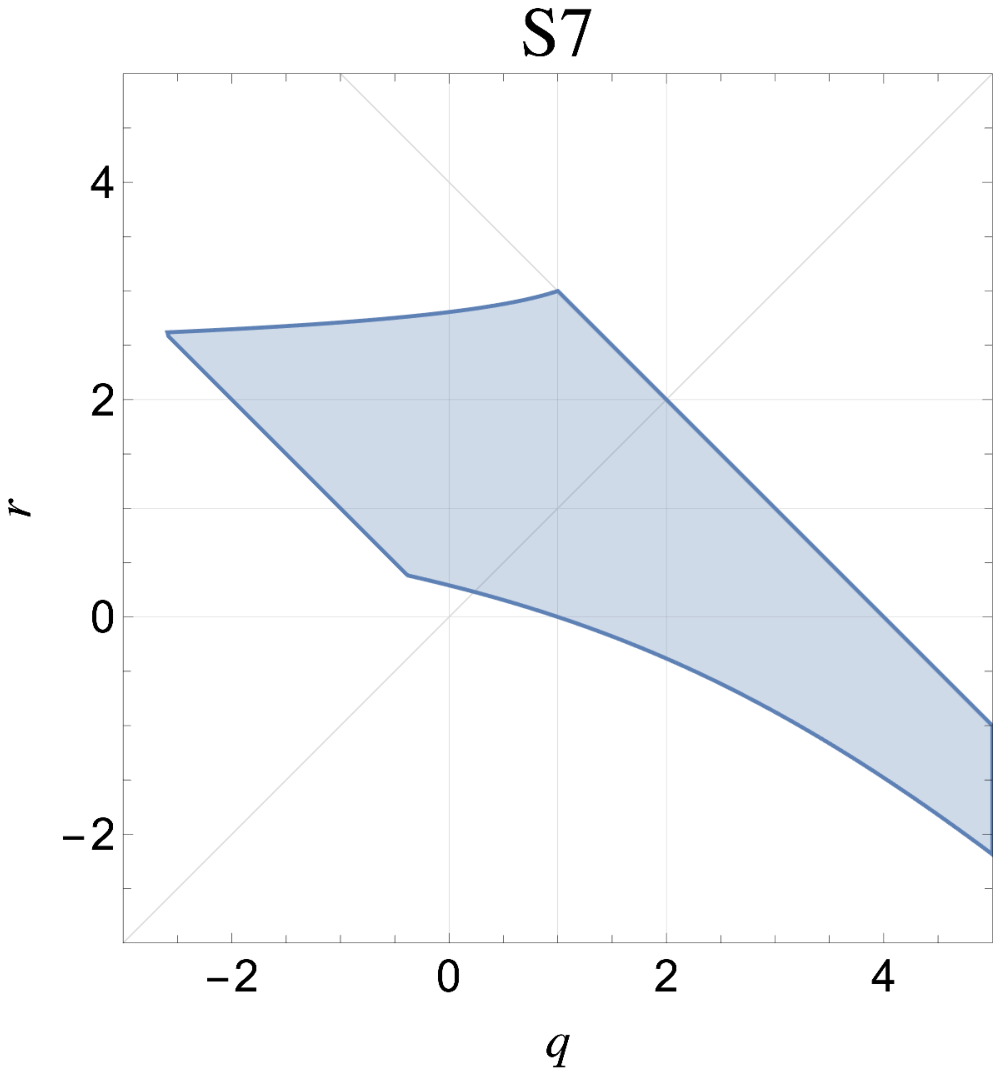}
\caption{Domains of definition of the solutions S4, S5 and S7.\label{fig:Sasymgen}}
\end{figure}

Then we have the following solutions with $\rho_4=1$ and domains of definition shown in figure \ref{fig:Sasymgen}:

\noindent$\bullet$ S4
\begin{align}
\begin{split}
&\rho_{0,1}=\rho_{3,1}=(\pm)\sqrt{\frac{q+r-4}{q^2-5q+5}}, \quad x_1= -\frac{q-4}{2}\rho_{0,1}, \\
& y_1=(\pm)\frac{1}{2} \sqrt{\frac{(4-q) \left(q^2-7 q+8\right)-(q-2)^2 (r-1)}{q^2-5 q+5}},\\
&\rho_{0,2}=0, \quad \rho_{3,2}=2 x_2 = (\pm)\sqrt{ 4-q-r}, \quad
y_2=(\pm)\frac{1}{2} \sqrt{q+r},
\end{split}
\end{align}
$\bullet$ S5
\begin{align}
\begin{split}
&\rho_{0,1}=\rho_{3,1}=(\pm)\sqrt{\frac{q+r-4}{q^2-5q+5}} , \quad \rho_{0,2}=(\pm) \sqrt{ \frac{q+r-4}{r-3}}\, \, , \, \,  \rho_{3,2}=-(r-3) \rho_{0,2},\\
&x_1=-\frac{q-4}{2}\rho_{0,1} , \quad y_1 =(\pm)\frac{1}{2}\sqrt{\frac{(4-q) \left(q^2-7 q+8\right)-(q-2)^2 (r-1)}{q^2-5q+5}}, \\
&  x_2=-\frac{r-4}{2}\rho_{0,2}, \quad
y_2=(\pm)\frac{1}{2}\sqrt{\frac{4-q(r-2)^2-r(r-4)^2}{r-3}}, \label{S5}
\end{split}
\end{align}
$\bullet$ S7
\begin{align}
\begin{split}
&\rho_{0,1}=0 , \quad \rho_{3,1}=2x_1 \quad \rho_{0,2}= (\pm)\sqrt{\frac{q+r-4}{r-3}}, \quad \rho_{3,2}=-(r-3)\rho_{0,2}, \\
& x_1 = (\pm)\frac{1}{2}\sqrt{4-q-r}, \quad  y_1=(\pm)\frac{1}{2}\sqrt{q+r},\\
&x_2=-\frac{r-4}{2}\rho_{0,2} , \quad y_2=(\pm)\frac{1}{2}\sqrt{\frac{4-q(r-2)^2-r(r-4)^2}{r-3}}.
\end{split}
\end{align}

Finally there are solutions defined only for isolated values of $q$ and $r$, of which we only list those defined for nonzero integer values of $q$ and $r$:

\noindent $\bullet$ V7 defined for $q=3$, $r=1$
\begin{align}
\begin{split}
&\rho_{0,1}=\rho_{0,2}=\rho_{3,1}=0 , \quad x_1=x_2 = \frac{1}{2}\rho_{3,2}=\pm\frac{1}{\sqrt{2}}\sqrt{1-\rho_4^2} , \quad\theta=(\pm)\frac{\pi}{2} ,\\
& y_1=y_2=(\pm)\frac{1}{\sqrt{2}}\sqrt{1+\rho_4^2} , \quad \rho_6=(\pm)\sqrt{1-\rho_4^2} , \quad \rho_4\in(0,1),
\end{split}
\end{align}
$\bullet$ V9 defined for $q=3$, $r=1$
\begin{align}
\begin{split}
&\rho_{0,1}=\rho_{3,1}=0, \quad \rho_{0,2} = 2 x_1 = \frac{2}{3}x_2 = \frac{1}{2}\rho_{3,2} = \pm \sqrt{1-\rho_4^2} , \;\; \theta=(\pm)\frac{\pi}{2},\\
& y_2=y_1= (\pm)\frac{1}{2}\sqrt{3+\rho_4^2} , \quad \rho_6=(\pm) \sqrt{1-\rho_4^2} , \quad \rho_4 \in (0,1),
\end{split}
\end{align}
$\bullet$ V11 defined for $q=r=2$
\begin{align}
\begin{split}
&\rho_{0,1}=\rho_{3,1}=\rho_{3,2}=x_1 = 2 x_2 =\pm\sqrt{1-\rho_4^2} , \quad\rho_{0,2}=0 , \quad \theta=(\pm)\frac{\pi}{2},\\
&y_1=(\pm)1 \, \, , \, \, y_2= (\pm)\frac{1}{2}\sqrt{3+\rho_4^2} \, \, , \, \, \rho_6=(\pm)\sqrt{1-\rho_4^2} \, \, , \, \, \rho_4\in(0,1),
\end{split}
\end{align}
$\bullet$ V22 defined for $q=3,r=1$
\begin{equation}
\begin{aligned}
&\rho_{0,1}=\rho_{0,2} = 2 x_1 = \frac{2}{3}x_2 = \frac{1}{2}\rho_{3,2} = \pm \sqrt{1-\rho_4^2}, \quad \rho_{3,1} = 0, \quad \theta = (\pm)\frac{\pi}{2}, \\
&y_1 = y_2 = (\pm)\frac{1}{2}\sqrt{3+\rho_4^2}, \quad \rho_6 = (\pm)\sqrt{1-\rho_4^2}, \quad \rho_4 \in (0, 1).
\end{aligned}
\end{equation}

\section{Scattering eigenstates}
\label{eigen}
Within the scattering description of the coupled $q$-state and $r$-state Potts models we consider the states
\begin{equation}
\psi_k = B_k \sum_{\substack{\gamma_1 = 1 \\ \gamma_1 \ne \alpha_1}}^{q} A_{\alpha_1\gamma_1}A_{\gamma_1\alpha_1} + C_k \sum_{\substack{\gamma_2 = 1 \\ \gamma_2 \ne \alpha_2}}^{r} A_{\alpha_2\gamma_2}A_{\gamma_2\alpha_2}\,,\hspace{1cm}k=1,2\,,
\label{eigenstates}
\end{equation}
which scatter into themselves through the phases $\Phi_k$ given by
\begin{equation}
\begin{aligned}
\Phi_1 &= \frac{1}{2}\Bigg [ (q-2)S_{2,1} + (r-2)S_{2,2} + S_{3,1} + S_{3,2} \\
& \qquad\qquad + \sqrt{\Big ( (q-2)S_{2,1}-(r-2)S_{2,2} + S_{3,1} -S_{3,2} \Big )^2+ 4(q-1)(r-1)S_4^2} \Bigg ]\,,
\end{aligned}
\end{equation}
\begin{equation}
\begin{aligned}
\Phi_2 &= \frac{1}{2}\Bigg [ (q-2)S_{2,1} + (r-2)S_{2,2} + S_{3,1} + S_{3,2} \\
& \qquad\qquad - \sqrt{\Big ( (q-2)S_{2,1}-(r-2)S_{2,2} + S_{3,1} -S_{3,2} \Big )^2+ 4(q-1)(r-1)S_4^2} \Bigg ]\,.
\end{aligned}
\end{equation}
The coefficients in (\ref{eigenstates}) are given by
\begin{equation}
\begin{aligned}
B_k &= S_{3,2} + (r-2)S_{2,2} - (r-1)S_4 - \Phi_k\,, \\
C_k &= S_{3,1} + (q-2)S_{2,1} - (q-1)S_4 - \Phi_k\,.
\end{aligned}
\end{equation}
In the decoupled case $S_4=0$ the phases $\Phi_1$ and $\Phi_2$ reduce to $(q-2)S_{2,1} + S_{3,1}$ and $(r-2)S_{2,2} + S_{3,2}$, respectively, as they should (recall (\ref{Phi})). As a generalization of (\ref{phaseS}), in the coupled case the relations
\begin{equation}
\Phi_k = e^{-2\pi i\, \Delta_{\eta_k}}
\label{Phi_k}
\end{equation}
give the conformal dimensions $\Delta_{\eta_1}$ and $\Delta_{\eta_2}$ of the fields which create the particles in the $q$-state sector and in the $r$-state sector, respectively. 

Notice that in the limit $r\to 1$ relevant for correlated percolation the phase $\Phi_1$ for the $q$-state sector becomes
\begin{equation}
\Phi_1|_{r=1}= (q-2)S_{2,1} + S_{3,1}\,,
\end{equation}
which is the value (\ref{Phi}) for the decoupled $q$-state model. As a consequence (\ref{Phi_k}) yields for $\Delta_{\eta_1}|_{r=1}$ the value of the decoupled model, in spite of the fact that the coefficients
\begin{align}
B_1|_{r=1} &= S_{3,2} - S_{2,2} - (q-2)S_{2,1} - S_{3,1}\,, \\
C_1|_{r=1} &= -(q-1)S_4
\end{align}
are both nonzero in the coupled case $S_4\neq 0$. Hence, $\Delta_{\eta_1}$ provides an example of a quantity that can be evaluated directly at $r=1$, where the degrees of freedom in the $r$-state sector play no role, since their number is strictly zero. Other quantities, on the other hand, will be determined by ratios in which both the numerator and the denominator vanish at $r=1$. The limit $r\to 1$ exists, but is determined by evaluation at $r=1+\epsilon$, where the auxiliary $r$-state degrees of freedom are present and contribute. The three-point connectivity of random percolation provides an exact illustration of this mechanism \cite{DV_3point}.


\end{document}